\documentstyle[12pt]{article}
\begin{document}
\newtheorem{lem}{Lemma}
\newsavebox{\uuunit}
\sbox{\uuunit}
    {\setlength{\unitlength}{0.825em}
     \begin{picture}(0.6,0.7)
        \thinlines
        \put(0,0){\line(1,0){0.5}}
        \put(0.15,0){\line(0,1){0.7}}
        \put(0.35,0){\line(0,1){0.8}}
       \multiput(0.3,0.8)(-0.04,-0.02){12}{\rule{0.5pt}{0.5pt}}
     \end {picture}}
\newcommand {\unity}{\mathord{\!\usebox{\uuunit}}}
\newcommand{\eentwee}{-1 \leftrightarrow 2}
\newcommand{\een}{{(1)}}
\newcommand{\twee}{{(2)}}
\newcommand{\drie}{{(3)}}
\newcommand{\half}{{\textstyle\frac{1}{2}}}
\newcommand  {\Rbar} {{\mbox{\rm$\mbox{I}\!\mbox{R}$}}}
\newcommand  {\Hbar} {{\mbox{\rm$\mbox{I}\!\mbox{H}$}}}
\newcommand {\Cbar}
    {\mathord{\setlength{\unitlength}{1em}
     \begin{picture}(0.6,0.7)(-0.1,0)
        \put(-0.1,0){\rm C}
        \thicklines
        \put(0.2,0.05){\line(0,1){0.55}}
     \end {picture}}}
\newcommand{\cL}{{\cal L}}
\newcommand{\cB}{{\cal B}}
\newcommand{\cM}{{\cal M}}
\newcommand{\cN}{{\cal N}}
\newcommand{\cZ}{{\cal Z}}
\newcommand{\cmap}{{$\bf c$ map}}
\newcommand{\rmap}{{$\bf r$ map}}
\newcommand{\crmap}{{${\bf c}{\scriptstyle\circ}{\bf r}$ map}}
\newcommand{\Ka}{{K\"ahler} }
\renewcommand{\theequation}{\thesection.\arabic{equation}}
\newcommand{\Al}{Alekseevski\v{\i}}
\newcommand{\eqn}[1]{(\ref{#1})}
\newcommand{\QED}{{\hspace*{\fill}\rule{2mm}{2mm}\linebreak}}

\begin{titlepage}
\begin{flushright} CERN-TH.6302/91\\ KUL-TF-91/43\\ THU-91/22
\end{flushright}
\vfill
\begin{center}
{\large\bf Special geometry, cubic polynomials and homogeneous
quaternionic spaces}   \\
\vskip 5.mm
{\bf B. de Wit$^1$}\\
Theory Division, CERN\\ CH-1211 Geneva 23, Switzerland\\[0.3cm]

{\bf A. Van Proeyen$^2$} \\
Instituut voor theoretische fysica
\\Universiteit Leuven,
B-3001 Leuven, Belgium
\end{center}
\vfill
\begin{center}
{\bf ABSTRACT}
\end{center}
\begin{quote}
The existing classification of homogeneous quaternionic spaces is
not complete. We study these spaces in the context of certain
$N=2$ supergravity theories, where dimensional reduction induces
a mapping between {\em special} real, K\"ahler
and quaternionic spaces. The geometry of the real spaces is
encoded in cubic polynomials, those of the K\"ahler and
quaternionic manifolds in homogeneous holomorphic functions of
second degree. We classify all cubic polynomials that have an
invariance group that acts transitively on the real manifold.
The corresponding K\"ahler and quaternionic manifolds are
then homogeneous. We find that they lead to a well-defined subset
of the normal
quaternionic spaces classified by \Al\ (and the corresponding
special K\"ahler spaces given by Cecotti), but there is a new
class of rank-3 spaces of quaternionic dimension larger than 3.
We also point out that some of the rank-4 \Al\ spaces were not
fully specified and correspond to a finite variety of
inequivalent spaces. A simpler version of the equation that
underlies the classification of this paper also emerges in the
context of $W_3$ algebras.
\vskip 2mm
\hrule width 5.cm
\vskip 1.mm
{\small\small
\noindent $^1$ On leave of absence from the Institute for
Theoretical Physics, Utrecht, The Netherlands; E-mail BDEWIT at
CRNVMA.CERN.CH or FYS.RUU.NL\\
\noindent $^2$ Onderzoeksleider, N.F.W.O. Belgium;
Bitnet FGBDA19 at BLEKUL11}
\normalsize
\end{quote}
\begin{flushleft} CERN-TH.6302/91 \\ KUL-TF-91/43\\ THU-91/22  \\
December, 1991.
\end{flushleft}
\end{titlepage}

\section{Introduction.}

Supersymmetric field theories in a variety of
space-time dimensions give rise to non-linear sigma models with a
restricted target-space geometry. There are many
examples in the literature where this phenomenon led to
surprising results, sometimes with interesting connections to
mathematics. Furthermore, the fact that some of these
supersymmetric theories in different space-time dimensions are
related by (supersymmetric) dimensional reduction offers a way of
connecting seemingly unrelated geometries.

In the context of this
paper $N=2$ supergravity is relevant. In five space-time
dimensions, one may consider the coupling of a certain number
(say $n-1$) of supersymmetric abelian vector multiplets. As was
shown some time ago \cite{GuSiTo}, these theories are
characterized by a cubic polynomial in $n$ (real) variables,
which gives rise to a non-linear sigma model corresponding to a
real ($n-1$)-dimensional space. Some of these polynomials
correspond to symmetric spaces and are
related to Jordan algebras. After dimensional reduction of this
theory, one finds $N=2$ supergravity in four space-time dimensions,
coupled to $n$ abelian vector multiplets. It is known that the
non-linear sigma models in the four-dimensional theory correspond
to K\"ahler spaces of complex dimension $n$, characterized by a
homogeneous holomorphic function of second degree, depending on
$n+1$ complex variables \cite{dWVP}. Such K\"ahler
manifolds are called {\em special} \cite{special}. Special
K\"ahler geometry is
relevant to string theory, where compactifications of type-II
superstrings on $(2,2)$
superconformal field theories with central charge $c=9$ lead to
$N=2$ supergravity coupled to vector multiplets. The massless
scalars of these vector multiplets play the role of coordinates of
the moduli space of the conformal theories, so
that the study of supergravity may thus lead to interesting
results for the moduli geometry of certain superconformal
theories \cite{Seiberg}. Because certain
(tree-level) results for compactifications of the heterotic string
on a $(2,2)$ superconformal system depend only on the choice of
the conformal theory, special geometry plays a role
for all string compactifications of this type, which include
those on
Calabi-Yau spaces. Indeed, this fact has been verified in several
studies where various aspects of this intriguing connection have
been explored [4-8].

After dimensional reduction of four-dimensional $N=2$
supergravity coupled to $n$ vector supermultiplets to three
space-time  dimensions, one finds a non-linear sigma model
corresponding to a
quaternionic manifold of quaternionic dimension $n+1$. In this
way one thus obtains a class of quaternionic manifolds
whose structure is encoded in the homogenous holomorphic function
of the special K\"ahler manifold. Hence there exists a map
between special K\"ahler manifolds of complex dimension $n$ and
certain quaternionic manifolds of quaternionic dimension
$n\!+\!1$, which in \cite{CecFerGir} was called the $\bf c$ map.
It was also shown that the $\bf c$ map plays an interesting
role in string theory. When compactifying IIA and IIB
strings on the same conformal theory, the resulting non-linear
sigma models consist of a product space of a K\"ahler manifold and
a quaternionic manifold. The latter is also special, in the sense
that it is characterized in terms of a homogeneous holomorphic
function.  When comparing the result of the compactification of
the IIA to that of the IIB string, it turns out that the two
manifolds are interchanged according to the action of the $\bf c$
map \cite{Seiberg,CecFerGir}.

Likewise one can introduce the \rmap, which yields for every real
space that couples to $d=5$ supergravity the corresponding \Ka
space that one finds upon reduction to four space-time
dimensions. The \rmap\ thus assigns a \Ka space of complex
dimension $n$ to a real space of dimension $n\!-\!1$.
Supersymmetry combined with dimensional reduction, which
preserves supersymmetry, are the essential ingredients in these
two maps.

We shall use the term "special
geometry" for both the real spaces originating from five-dimensional
supergravity, the K\"ahler spaces originating from
four-dimensional supergravity and the quaternionic spaces
that are in the image of the $\bf c$ map\footnote{
In the literature, special K\"ahler spaces  were sometimes called
K\"ahler spaces of restricted type; the special quaternionic
spaces were also called dual-quaternionic spaces.}.
It should be clear that the inverse $\bf r$  and $\bf c$ maps
are not always defined\ as there are spaces that couple to
supergravity, but the corresponding supergravity theory does not
necessarily originate from a higher-dimensional theory. When the
coupling of a certain
space to supergravity is not unique, the result of
the maps will depend on the type of coupling as, for instance,
characterized by the way in which the
subgroup  of the sigma model isometries that can be extended to a
symmetry of the full supergravity action, is realized. In four
space-time dimensions these invariances
usually act on the field strengths of the abelian vector fields,
and not on the fields themselves, so that they only leave the
equations of motion and not the action invariant. These
transformations, called duality transformations,
constitute a subgroup of  $Sp(2n\!+\!2,\Rbar)$. The complex
nature of the \Ka manifolds is thus related to the complex nature
of the (anti-)selfdual Minkowskian field strengths
\cite{dual,dWVP}.
In five space-time dimensions, we are dealing with a real
manifold, so that the transformations are realized directly on
the vector fields, whereas in three dimensions the vector fields
are converted into scalar fields (the relation of all these
symmetries upon dimensional reduction will be discussed in
\cite{dWVPVan}; see also \cite{sbstring}).
Under these maps the dimensionality of the manifold increases.

For homogeneous spaces the isometries act transitively on
the manifold so that every two points are related by an element
of the isometry group. The orbit swept out by the action of the isometry
group $G$ from any given point is (locally) isomorphic to the
coset space $G/H$, where $H$ is the isotropy group of that point.
For non-compact homogeneous spaces where $H$ is the maximal compact
subgroup of $G$, there exists a solvable subgroup that acts
transitively, whose dimension is equal to the dimension of the
space. Such spaces are called {\em normal}. It implies that there
exists a
solvable algebra $s$ such that $\frac{G}{H}=e^s$. The
construction of this algebra follows from the
Iwasawa decomposition of the algebra $g=h+s$ (see e.g. \cite{Helg}). The
dimension of the Cartan subalgebra of $s$ equals the {\it rank} of the
homogeneous space. It will turn out that the rank of the symmetry algebra
and of its solvable subalgebra increase by one unit under the
$\bf c$ and \rmap s.
In the context of this paper the following considerations are important. If
the result of the $\bf c$ map is a homogeneous quaternionic
space, then the duality invariance (the symmetry of the
scalar-vector sector of the theory) of the original theory must
act transitively on the corresponding
manifold parametrized by the scalar fields. The proof of this
result, which applies also to the $\bf r$ map, is given in
\cite{dWVPVan}. Also the converse is true: if the
vector-scalar symmetries act transitively on the manifold
parametrized by the scalars, then one can show that the symmetry
group after dimensional reduction gives rise to additional
symmetries, which leave the original scalar fields invariant but
act transitively on the new scalar fields. In this respect it is
important that the process of dimensional reduction always entails new
symmetries whose number is larger than or equal to the
number of new coordinates.

The above results show that homogenous quaternionic spaces that
are in the image of the $\bf c$ map correspond to special homogeneous
K\"ahler spaces. On the other hand, special
homogeneous K\"ahler spaces give rise to special homogeneous
quaternionic spaces, provided that the scalar-vector symmetry
transformations act transitively on the K\"ahler manifold.
Likewise, such
K\"ahler spaces that are themselves in the image of the $\bf r$ map
correspond to special homogeneous real spaces. Again, special
homogeneous real spaces give rise
to homogeneous K\"ahler spaces, provided that the vector-scalar
symmetries act transitively on the real manifold.

In this connection \Al's classification of homogeneous
quaternionic spaces \cite{Aleks} is relevant, as was first
pointed out in \cite{CecFerGir}.
In \cite{Aleks} it was conjectured that the homogeneous
quaternionic  spaces consist of compact symmetric
quaternionic spaces and  (non-compact) normal quaternionic spaces.
Normal quaternionic spaces are quaternionic spaces that admit a
transitive completely solvable group of motions.
According to \Al\ there are two different
types of normal quaternionic spaces characterized by their
so-called canonical quaternionic subalgebra. The first type has as
canonical subalgebra $C_1^1$, the solvable algebra corresponding to
$Sp(1,1)/(Sp(1)\otimes Sp(1))$, and corresponds
to the quaternionic projective spaces $Sp(m,1)/(Sp(m)\otimes Sp(1))$.
These spaces are {\em not} in the image of the $\bf c$ map.
The second type has a canonical subalgebra $A_1^1$, the solvable subalgebra
of $SU(2,1)/(SU(2)\otimes U(1))$. Denoting the  dimension of the normal
quaternionic algebra as $4(n\!+\!1)$, the
structure of the solvable algebra is such that it always contains a normal
K\"ahler algebra ${\cal W}^{\rm s}$ of dimension $2n$,  whose action on
the remaining part of the algebra naturally defines a
$(2n\!+\!2)$-dimensional representation corresponding to a
solvable subgroup of $Sp(2n+2, \Rbar)$.\footnote{This
representation thus acts on $2n\!+\!2$ of the
generators. The two remaining generators of
in the quaternionic algebra, $e_0$ and $e_+$, are inert under
${\cal W}^{\rm s}$. The sum of the Cartan
subalgebra of ${\cal W}^{\rm s}$ and $e_0$ constitutes the
Cartan subalgebra of the quaternionic algebra, whose rank is thus
1 higher than that of the K\"ahler
algebra. The weight of the K\"ahler algebra under $e_0$ is thus zero,
while the weight of the generators that constitute the
$(2n+2)$-dimensional representation is 1/2 times the weight of
$e_+$.}
Therefore each
normal quaternionic space of this type defines the basic ingredients of
a special normal K\"ahler space, encoded in its solvable
transitive group of duality transformations.
\Al's analysis thus strongly indicates that the corresponding
$N\!=\!2$ supergravity theory should exist, so that under
the $\bf c$ map one will recover the original normal quaternionic
space. To establish the existence of  the supergravity theory,
one must prove that a corresponding holomorphic function
$F(X)$ exists that allows for these duality transformations. This
program was
carried out by Cecotti \cite{Cecotti}, who explicitly constructed
the function $F(X)$ corresponding to each of the normal quaternionic
spaces with canonical subalgebra $A^1_1$ that appears in the
classification of \Al. With the
exception of the so-called minimal coupling, where $F(X)$ is a quadratic
polynomial, all the \Ka spaces are in the image of the \rmap. The
corresponding special K\"ahler manifolds were
denoted by $H(p,q)$ and $K(p,q)$. Under the $\bf c$ map, they
lead to the normal quaternionic manifolds $V(p,q)$ and $W(p,q)$
defined in \cite{Aleks}. If \Al's classification is complete,
there can be no other special K\"ahler
spaces with solvable transitive duality transformations.

In this paper we start at the other end and derive a
classification of all homogeneous
quaternionic spaces that are in the image of the \crmap.
The analysis can be performed completely at the level of the
special real spaces, and amounts to classifying all the cubic polynomials
whose invariance group acts transitively
on the corresponding special real spaces. This invariance group
leaves the full $d=5$ supergravity Lagrangian invariant. The
corresponding  real spaces are obviously homogeneous, but because
of the results quoted above, so are the corresponding
\Ka and quaternionic spaces that emerge under the action of the \rmap\
and the \crmap. When comparing the result to the
classification of \Al\ (and the corresponding one of Cecotti) we
find that their classification is incomplete!

The cubic functions that are classified in this paper, are
parametrized by
\begin{equation}
C(h) = d_{ABC}\,h^A\,h^B\,h^C \ . \qquad (A,B,C = 1,
\ldots, n) \label{Cpoly}
\end{equation}
The corresponding sigma model, which is contained in the
five-dimensional supergravity Lagrangian \cite{GuSiTo}, is
defined by the Lagrangian
\begin{equation}
{\cal L} = -{\textstyle{3\over 2}} d_{ABC} \,h^A\,\partial_\mu
h^B\, \partial^\mu h^C \ ,   \label{sigma}
\end{equation}
where the scalar fields $h^A$ are restricted by $C(h)=1$, so that
the sigma model corresponds to a $(n\!-\!1)$-dimensional real space.
Linear redefinitions of the fields $h^A$ that leave $C(h)$
invariant constitute invariances of the full $N\!=\!2$
supergravity Lagrangian. However, it is not excluded
that the sigma model Lagrangian \eqn{sigma} has additional symmetries,
which cannot be extended to symmetries of the full
supersymmetric Lagrangian.
The polynomial $C(h)$ is left invariant by linear transformations
of the fields $h^A$, whose infinitesimal form is parametrized by
matrices $B^A_{\;B}$,
\begin{equation}
\delta h^A = B^A_{\;B}\,h^B \ ,   \label{Btrans}
\end{equation}
restricted by the condition
\begin{equation}
B^D_{\,(A}\,d_{BC)D} = 0 . \label{Binv}
\end{equation}

As explained above, our aim is to determine all tensors $d_{ABC}$
whose invariance group acts transitively on the manifold defined
by (\ref{sigma}). To analyze this question we first
redefine the scalar fields in some reference
point where the metric associated with the sigma model has
positive signature\footnote{Positive signature is required to ensure
a positive-definite Hilbert space of physical states. The
necessary and sufficient condition for this is that the variables
$h^A$ are restricted to a domain where
\[
\left(3d_{ACD}\,d_{BEF}-2d_{ABC}\,d_{DEF}\right) h^Ch^Dh^Eh^F
\]
is a positive definite matrix.}.
One may
choose this reference point equal to $h^A= (1, 0,\ldots,0)$. In
that  case the coefficients $d_{ABC}$ can be redefined according
to the so-called canonical parametrization
\begin{equation}
d_{11a} =0,\quad d_{1ab} =-\half \,d_{111}\,\delta_{ab}\ . \qquad (a,b = 2,
\ldots, n) \label{canpar}
\end{equation}
with $d_{111}>0$. To preserve this parametrization only
orthogonal redefinitions of the fields $h^a$ are allowed.

The condition \eqn{Binv} that the $C(h)$ be invariant
is then analyzed in the canonical parametrization. Putting
$d_{111}=1$ for convenience, \eqn{Binv} implies
that $B^A{}_{\!B}$ takes the following form (see \cite{BEC} where
the corresponding K\"ahler spaces were analyzed)
\begin{equation}
B^1{}_1 = 0\ ,\quad   B^a{}_1 =
B^1{}_a\ , \quad    B^a{}_b = B^c{}_1 \,
d_{abc}  + A_{ab}
\label{cBpar}
\end{equation}
where $A_{ab}$ is an antisymmetric matrix with $a,b,\ldots = 2,
\ldots,n$. This matrix is subject to the condition
\begin{equation}
\Gamma_{abcd}\,  B^d{}_1 =  d_{d(ab}\, A_{c)d} ,
\end{equation}
where
\begin{equation}
\Gamma _{abcd}\equiv d_{e(ab}\,d_{cd)e}-{\textstyle\half} \,
\delta_{(ab}\,\delta_{cd)}\ .
\end{equation}

Now we observe that transformations associated with
the matrices $A_{ab}$ that are independent of the parameters
$B^a{}_1$, leave the canonical reference point invariant and
thus correspond to the isotropy group. Hence we are left with the
requirement that the symmetry group should contain $n-1$
independent parameters $B^a{}_1$. Writing $A_{ab} =
B^c{}_1 \,A_{ab;c}\,$, where $A_{ab;c}$ is antisymmetric in its
first two indices, this leads to the equation
\begin{equation}
\Gamma _{abcd}=   D_{abc;d}\ , \label{GamD1}
\end{equation}
where $\Gamma _{abcd}$ was defined above and
\begin{equation}
D_{abc;d} = d_{e(ab}\,A_{c)e;d} \ .
\end{equation}
{}From the above results, it is clear that
the homogeneous real manifold corresponding to (\ref{sigma}) is
locally isomorphic to $G/H$, where $G$ is the invariance group of
the tensor $d_{ABC}$ and $H$ is the orthogonal invariance group
of the tensor $d_{abc}$.

{}From the arguments given earlier it follows that there is a
corresponding analysis for the special \Ka and quaternionic
spaces that follow from the real spaces that we introduced above.
One may thus consider the \Ka spaces and require that the
symmetry group of the $d=4$ supergravity Lagrangian acts
transitively on the space.
The cubic polynomial $C(h)$ is directly
related to the holomorphic function $F(X)$, which encodes the
information of the special K\"ahler manifolds that follow from
the real manifolds by the $\bf r$ map. It reads
\begin{equation}
F(X) = id_{ABC}{X^A\,X^B\,X^C\over X^0} \ ,    \label{Fd}
\end{equation}
where $X^0$ and $X^A$ are complex variables. The K\"ahler manifold is
only $n$-dimensional because two points $(X^0,X^A)$ that are
related by multiplication with an arbitrary complex number are
identified. The \rmap\ thus introduces $n\!+\!1$ new coordinates,
but at the same time it leads to at least $n\!+\!1$ additional
symmetries \cite{dWVPVan,sbstring} so that the analysis proceeds
along the same steps. Similarly, for quaternionic manifolds, the
requirement of transitivity rests upon the same analysis as
presented for the real manifolds\footnote{The isometries for
the special quaternionic spaces were analyzed in \cite{dWVP2}.}
Therefore there is no need for going into further details.

A special case of \eqn{GamD1} (namely $\Gamma_{abcd}=0$) was
analyzed in \cite{GuSiTo} in the context of Jordan algebras and
in \cite{BEC} for the special \Ka spaces. The connection with
Jordan algebras arose because the \eqn{GamD1} is equivalent to
the condition that the torsion tensor associated with the special
real space is covariantly constant. In that case the real space
is symmetric (but this does not exhaust the special symmetric
spaces). Likewise the corresponding \Ka and quaternionic spaces
that one obtains by means of the \rmap\ and the \crmap\ are
symmetric (and in this case there are no other (special)
symmetric spaces).  Surprisingly, the equation $\Gamma_{abcd}=0$
emerges also in a different context, namely that of
$W_3$ algebras \cite{Zam}, where it corresponds to the condition that
ensures that the higher-spin invariances of a
two-dimensional conformal field theory can be
consistently truncated to
the energy-momentum tensor and a spin-3 charge \cite{Wred}.

In view of these
and possible other applications of \eqn{GamD1}, we shall keep the
analysis of \eqn{GamD1} self-contained without using the
connection with the special geometries. The central result of
this paper, namely the
classification of the tensors $d_{abc}$ that satisfy \eqn{GamD1},
is presented in section~\ref{proofd}. The reader who is
only interested
in the results can skip this section as well as section 3, where
we rewrite the results for $d$ in a simpler form and present the
solutions for the tensors $A$ in \eqn{GamD1}. The final result
for the cubic polynomial $C(h)$ is given in section 4. It can
be expressed as follows (not in the canonical
parametrization). First we decompose the coordinates $h^A$ into
$h^1$, $h^2$, $h^\mu$ and $h^m$, where the range of the indices
$\mu$ and $m$  is equal to $q+1$ and $r$, respectively. Hence we have
\begin{equation}
n= 3+q+r ,
\end{equation}
so that $n\geq 2$. Then $C(h)$ can be written as
\begin{equation}
C(h) = 3\Big\{ h^1\,
\big(h^2\big)^2 -h^1\,\big(h^\mu\big)^2 -h^2\,\big(h^m\big)^2
+\gamma_{\mu mn}\,h^\mu\, h^m\,h^n\Big\} \ ,\label{genC1}
\end{equation}
where the coefficients $\gamma_{\mu mn}$ are the generators of a
$(q\!+\!1)$-dimensional real Clifford algebra with positive
signature.

Further results and and implications are given in section~\ref{conclusions}.

\section{Classification.}          \label{proofd}
\setcounter{equation}{0}
In this section we study
\begin{equation}
\Gamma _{abcd}=   D_{abc;d}\ , \label{GamD}
\end{equation}
where the indices $a, b,\dots$ take $n-1$ values $2, \ldots, n$,
\begin{eqnarray}
\Gamma _{abcd}&=& d_{e(ab}\,d_{cd)e}
-\half \delta _{(ab}\,\delta_{cd)}\ , \\
D_{abc;d} &=& d_{e(ab}\,A_{c)e;d} \ ,
\end{eqnarray}
and $A_{ab;c}$ is a tensor that is antisymmetric in its first
two indices. Observe that $D_{abc;d}$ is only manifestly symmetric
in three indices; full symmetry is only obtained after imposing
\eqn{GamD}. The tensors $d_{abc}$ are symmetric and are concisely
summarized by the cubic polynomial,
\begin{equation}
{\cal Y}(x) = d_{abc}\,x_a\,x_b\,x_c\ .   \label{cubic}
\end{equation}

We will now give a complete classification of
the tensors $d_{abc}$ that satisfy \eqn{GamD} up to orthogonal
redefinitions. Obviously, the tensors
$A_{ab;c}$ can only be determined up to the generators of
orthogonal transformations
that leave $d_{abc}$, and thus the function ${\cal Y}(x)$
invariant.  The analysis is done in two
steps. First we show that after a suitable
$O(n-1)$ rotation, it is always possible to
bring the tensors $d_{abc}$ into a form such that
\begin{eqnarray}
d_{22a} &=&  \frac{1}{\sqrt 2} \, \delta_{a2}\ ,
\nonumber \\
\Gamma_{222a} &=& A_{a2;2} = 0\ . \label{step1}
\end{eqnarray}
The second step is then to bring the $d_{abc}$ coefficients
in a form where $d_{2ab}$ is diagonal for general $a$
and $b$ and examine the consequences of \eqn{GamD}.

Let us start by using $O(n-1)$ transformations to define
a "2" direction (which will not necessarily coincide with the
"2" direction chosen in \eqn{step1}) such that
\begin{equation}
d_{abb}=\lambda \,\delta_{a2}\; .\label{deflambda}
\end{equation}
A contraction of \eqn{GamD} over two indices then implies that the
following three tensors must be equal,
\begin{eqnarray}
3\Gamma _{abcc}&=&2 d_{acd}\,d_{bcd}
+\lambda\, d_{2ab} -\half (n+1)\,\delta _{ab} \;,\nonumber\\
3D_{cca;b}&=& \lambda\, A_{a2;b}\; , \nonumber\\
3D_{abc;c}&=& 2 d_{ec(a}\, A_{b)e;c}+ d_{abc}\,A_{dc;d}\;.
\label{cGamD}
\end{eqnarray}

Now we distinguish between three different cases, denoted by I, II
and III, which will play
a role throughout this analysis.

In case I we have
\begin{equation}
\lambda= 0 \Longleftrightarrow d_{abb} = 0\;.
\end{equation}
According to \eqn{cGamD} we then have
\begin{equation}
\Gamma_{ccab}= D_{cca;b}=D_{abc;c} = 0\;. \label{Deq}
\end{equation}
Using a notation where $d_a$ and $A_a$ are $(n-1)\times(n-1)$
matrices defined by
$(d_a)_{bc}\equiv d_{abc}$ and $(A_a)_{bc} \equiv A_{bc;a}$,
the first equation \eqn{Deq} reads
\begin{equation}
\langle d_a\, d_b\rangle = \textstyle{1\over 4}(n+1)\,
\delta _{ab} \;,
\end{equation}
where $\langle A \rangle$ denotes the trace of a matrix $A$.
Making use of this result we contract the tensors appearing in
\eqn{GamD} with $d_{cdf}$, leading to
\begin{eqnarray}
3\Gamma _{abcd}\,d_{cdf}&=&\textstyle{1\over 4}(n-3)\,d_{abf}
+2 \langle d_a\, d_f\,d_b\rangle \ , \nonumber\\
3D_{acd;b}\, d_{cdf}&=&\textstyle{1\over 4} (n+1)\,
A_{af;b}+2\langle d_a\,d_f\,A_b \rangle\ .  \label{l0}
\end{eqnarray}
According to \eqn{GamD} these two tensors should be equal. However,
the first one is symmetric and the second one antisymmetric in $a$
and $f$. Therefore they should vanish separately.
Combining the above results, we derive
\begin{equation}
\Gamma _{abcd}\,\Gamma _{abce}=
\Gamma _{abcd} \,D_{abc;e}= -\half D_{aad;e} =0.
\end{equation}
For case I we therefore obtain
\begin{equation}
\Gamma _{abcd}=D_{abc;d} = d_{abb}=0 .
\end{equation}
These are the equations that were analyzed in
the appendix of \cite{BEC}. The first part of
this analysis coincides with the one
that we are about to present for case II and III in the limit
where the $A_{ab;c}$ tensors are put to zero or coincide with
generators of the
$O(n-1)$ subgroup that is left invariant by $d_{abc}$.
A minor complication
is that the "2" direction is not yet defined for case I,
in view of the fact that $d_{abb}=0$. However, the analysis only
requires that $d_{222}\not=0$.

Hence we proceed to case II and III where $\lambda\not=0$.
Therefore we know from  \eqn{cGamD} that
$A_{2a;b}$ is symmetric in $a$ and $b$.
{}From this it follows that $A_{ab;c}=0$
whenever two of its indices are equal to 2, which leads to
$D_{222;2}=0$.

In case II we assume that $d_{22i}= 0$, where $i=3,\ldots ,n$.
Using $\Gamma_{2222}= D_{222;2}=0$ one finds that
\begin{equation}
(d_{222})^2 = \half .
\end{equation}
As we can choose the sign of $d_{222}$ at will, we thus find that
case II also leads to \eqn{step1}.

In case III we assume that not all $d_{22i}$ vanish. Diagonalizing
the symmetric matrix $d_{2ij}-A_{2i;j}$ gives
\begin{equation}
A_{2i;j}=d_{2ij}-\lambda _i\,\delta _{ij}.  \label{diagdA}
\end{equation}
Then $\Gamma_{222i}= D_{222;i}$ yields
\begin{equation}
d_{22i}\left( \alpha +\lambda _i\right) =0 , \nonumber
\end{equation}
where we used the notation $\alpha \equiv d_{222}$.
For those values of $i$ for which $d_{22i}\neq 0$, we have
the same eigenvalue $\lambda _i=-\alpha$.
Hence, by means of a rotation, we can define a "3"
direction such that
\begin{equation}
d_{22i}=\beta \, \delta_{i3}\   , \qquad \lambda_3 =-\alpha ,
\end{equation}
where $\beta \not=0$ (otherwise we would be dealing with case II).
Using $\Gamma_{222i} =D_{22i;2}$ gives
\begin{equation}
d_{233}=-\alpha \ ,\qquad
A_{\alpha 3;2}=3d_{23\alpha } \ ,
\end{equation}
with $\alpha =4,\ldots ,n$.
Hence we also have
\begin{equation}
A_{23;3} = d_{233}-\lambda_3 = 0\ .
\end{equation}
Then from $\Gamma_{2222} = D_{222;2} = 0$, one derives
\begin{equation}
\alpha ^2+\beta ^2=\half\ .\label{2222}
\end{equation}
Now we analyze \eqn{GamD} with indices
$(2233)$. First $D_{332;2}=D_{223;3}$ takes the form
\begin{equation}
-2\left( d_{23\alpha }\right) ^2
={\textstyle\frac{2}{3}}\left( d_{23\alpha }\right) ^2 \ .
\end{equation}
Combining the above equations gives
\begin{equation}
d_{23\alpha }=A_{\alpha 3;2}=A_{23;\alpha }=A_{2\alpha ;3}=0\ .
\end{equation}
This implies that also $D_{223;\alpha }=0$ and thus
\begin{equation}
3D_{22\alpha ;3}=\beta\,A_{\alpha 3;3}=0.
\end{equation}
Hence the tensor $A_{ab;c}$ vanishes whenever two of its indices
are equal to 2 or 3. Moreover we have $A_{2\alpha;\beta} =
A_{2\beta;\alpha}$.

Subsequently we deduce from $D_{332;2}=D_{223;3}=0$ that
$\Gamma _{2233}=0$. Combining this with
\eqn{2222} shows that
\begin{equation}
d_{333}=-\beta \ .
\end{equation}
Then $\Gamma _{223\alpha }=0$ gives
\begin{equation}
d_{33\alpha }=0\ .
\end{equation}

Hence our results for the $d$ coefficients take the form
\begin{eqnarray}
&&d_{222}=\alpha \ ,\ d_{223}=\beta \ ,\ d_{233}=-\alpha \ ,\
d_{333}=-\beta \nonumber\\
&&\alpha ^2+\beta ^2=\half \ ,\nonumber\\
&&d_{22\alpha }=d_{23\alpha }=d_{33\alpha }=0\nonumber\\
&&d_{2bb}= d_{2\beta\beta} =\lambda \not= 0\ ,\nonumber \\
&& d_{3bb} = d_{3\beta\beta} = 0\ , \ d_{\alpha bb}=
d_{\alpha \beta \beta }= 0\ .
\end{eqnarray}
Now we may perform an $O(2)$ transformation in the
$(2,3)$ space such that the new coefficient
$d_{223}$ vanishes. In terms of the cubic function ${\cal Y}(x)$
this transformation corresponds to an orthogonal redefinition of
$x_2$ and $x_3$,
\begin{eqnarray}
x'_2 &=& x_2\cos \phi  \pm x_3\sin\phi\ , \nonumber\\
x'_3 &=& -x_2\sin \phi  \pm x_3\cos\phi \ .
\end{eqnarray}
Using \eqn{2222} and defining $\alpha
=\frac{1}{\sqrt{2}}\cos \theta $ and $\beta
=\frac{1}{\sqrt{2}}\sin \theta $, we obtain a one-parameter family
of coefficients
\begin{equation}
\alpha '=\frac{1}{\sqrt{2}} \cos (\pm\theta -3\phi )\ ;\qquad
\beta '=\frac{1}{\sqrt{2}} \sin (\pm\theta -3\phi ). \label{phiab}
\end{equation}
We can thus choose a parametrization such that
\begin{eqnarray}
&& d_{222}=\frac{1}{\sqrt{2}}\ , \qquad
d_{233}=-\frac{1}{\sqrt{2}}\ , \nonumber \\
&& d_{223}= d_{22\alpha} = d_{333} = d_{23\alpha} = d_{33\alpha}
= d_{\alpha\beta\beta}=0  \ ,
\end{eqnarray}
so that case III also allows the parametrization \eqn{step1}.
Observe that
after this redefinition $d_{abb}$ may only differ from zero for
$a=2$ or 3. Hence
\begin{equation}
d_{abb}= \lambda_2\,\delta_{a2} + \lambda_3\,\delta_{a3} \ .
\end{equation}
Case I is now characterized by $\lambda_2=\lambda_3=0$, case II
by $\lambda_2\not=0\ , \lambda_3=0$, and case III by $\lambda_3\not=
0$.

Note, however, that the angle $\phi$ in \eqn{phiab} is not uniquely
determined.  There are 6 solutions.  This means that there is still the
possibility of redefining $x_2$ and $x_3$, such that we remain
within the parametrization \eqn{step1}. Those redefinitions
consist of products of reflections,
\begin{equation}
x_3 \rightarrow -x_3 \label{ch3m}  \ ,
\end{equation}
and  $2\pi /3$ rotations,
\begin{eqnarray}
x_2 &\rightarrow &-\textstyle\frac{1}{2}\, x_2
+ \textstyle\frac{1}{2}\sqrt{3}\,x_3\ ,   \nonumber\\
x_3 &\rightarrow &-\textstyle{1\over2}\sqrt{3}\, x_2
- \textstyle\frac{1}{2} \,x_3\ .   \label{ch23}
\end{eqnarray}
These replacements do not change the part of $\cal Y$ that is
quadratic or cubic in $x_2$ and $x_3$,
\begin{equation}
{\cal Y}(x) =\frac{1}{\sqrt{2}}
\left( x_2^{\,3}-3x_2\,x_3^{\,2}\right) +\ldots .
\end{equation}
Later we shall see that the
above redefinitions allow one to rewrite some of
the solutions belonging to case II
into those belonging to case III.
\QED

This concludes the proof of \eqn{step1}. The second step in the
classification starts by diagonalizing
$d_{2ab}$ for all $a$ and $b$ (this
is consistent with \eqn{step1}). Hence we adjust the
frame of reference, such that
\begin{equation}
d_{2ij }= \mu _i \, \delta _{ij} \ ,
\end{equation}
where we recall that $i,j=3,\ldots, n$.

Now we consider \eqn{GamD} with indices $(22ij)$, according
to which the following three tensors should be equal,
\begin{eqnarray}
3\Gamma _{22ij }&=&\Big( \frac{1}{\sqrt{2}}
\,\mu _i +2\mu _i^2-\half\Big)
\delta _{ij} \ , \nonumber\\
3D_{22i;j}&=&\Big( -\frac{1}{\sqrt{2}}
+2\mu _i\Big) A_{2i;j}\  , \nonumber\\
3D_{ij2;2}&=&(\mu _j -\mu _i )\, A_{ij;2}\ .
\end{eqnarray}
As the last tensor vanishes for
$i=j$, while the first one takes its non-zero values in that case,
the three tensors should vanish separately.  The vanishing of
the first one implies that
$\mu _i$ can only take two possible values, $-\frac{1}{\sqrt{2}}$ or
$\frac{1}{2\sqrt{2}}$.  Therefore it is convenient to
split the indices $i$ according to
these values into indices $\mu, \nu, \ldots$ and $m, n, \ldots$ such
that
\begin{equation}
\mu _\mu =-\frac{1}{\sqrt{2}}\ ,\qquad \mu _m=\frac{1}{2\sqrt{2}}\ .
\end{equation}
Furthermore we obtain
\begin{equation}
A_{\mu m;2}= A_{2\mu;\nu  }=A_{2\mu ;m}=0.\label{A20}
\end{equation}
It is clear that the special
index value $i=3$ that occurred in the analysis of
case III, is contained in the index set labeled by
$\mu, \nu, \ldots$.

The next step is the analysis of \eqn{GamD} with indices
$(2ijk)$. The corresponding tensors are
\begin{eqnarray}
3\Gamma _{2ijk}&=&d_{ijk}\left( \mu _i+\mu _j+\mu _k\right)\ ,
\nonumber\\
3D_{ijk;2} &=&3d_{l(ij}A_{k)l;2}\ , \nonumber\\
3D_{2ij;k} &=&d_{lij}A_{2l;k} + (\mu _i-\mu _j)\,
A_{ji;k} \ .\label{2ijk}
\end{eqnarray}
Using \eqn{A20} it follows that $d_{\mu\nu\rho}\,D_{\mu\nu\rho;2}
= d_{\mu\nu m}\,D_{\mu\nu m;2} = d_{mnp}\,D_{mnp;2} = 0$ by
virtue of the antisymmetry of the coefficients $A_{ij;2}$.
Therefore the tensor $\Gamma_{2ijk}$ should vanish when contracted with
these $d$ coefficients. As $\Gamma_{2ijk}$ is itself
proportional to the
$d$ coefficients, it follows that certain components
should vanish, i.e.,
\begin{equation}
d_{\mu \nu \rho }=d_{\mu\nu m} = d_{mnp}=0\ .
\end{equation}
As $\Gamma_{2\mu mn}$ already vanishes by virtue of the
fact that $\mu_\mu+\mu_m+\mu_n =0$, we have thus established that
all components of the
$\Gamma$ tensor with one or more indices equal to 2 now vanish.
Most of the components of $D_{ijk;2}$ and $D_{2ij;k}$ now
vanish identically. The equation $ D_{\mu mn;2} = 0$ implies
that the $d$ tensors should be left invariant by orthogonal
transformations characterized by
the $A_{ij;2}$. The latter can be put to zero and do
not restrict the $d$-coefficients. Furthermore, there is
\begin{equation}
D_{2\mu m;i} =0\ \Longleftrightarrow  \
A_{m\mu ;i } =  \textstyle{2\over 3}\sqrt{2} \,
d_{\mu mn}A_{2n;i}\ .\label{condmi}
\end{equation}

The only components of $\Gamma_{ijkl}$ that do not vanish
identically at this point, are
\begin{eqnarray}
\Gamma _{\mu \nu mn}&=&\textstyle{\frac{2}{3}}
d_{mp(\mu }\,d_{\nu) np}
-\textstyle{\frac{1}{4}}\delta_{\mu \nu }\,
\delta _{mn}\ ,\label{defGmu}\\
\Gamma _{mnpq}&=&d_{\mu (mn}\,d_{pq)\mu }
-\textstyle{\frac{3}{8}} \delta _{(mn}\, \delta_{pq)}\ .
\label{defGm}\end{eqnarray}
According to \eqn{GamD}, they should satisfy
\begin{eqnarray}
\Gamma _{\mu \nu mn}&=&D_{\mu \nu m;n}=- \Gamma _{\mu
\nu mp}\,H_{pn}\ ,\label{GmuH} \\
\Gamma _{\mu \nu mn}&=& D_{mn\mu ;\nu} =\textstyle{2\over3}\,
d_{\mu q(m}\,A_{n)q;\nu }+\textstyle{1\over3}\,d_{\rho mn}\,
A_{\mu \rho  ;\nu }\ , \label{GmuA} \\
\Gamma _{mnpq}&=&D_{mnp;q} =\Gamma_{mnpr}\,H_{rq}\ ,
\label{GmH}
\end{eqnarray}
where we made use of \eqn{condmi} and
defined $H_{mn}\equiv {2\over 3}\sqrt 2 A_{2m;n}$.

Contractions of the above equations will give useful
information.
Denoting the range of the indices $\mu $ by $q+1$, and the range
of the indices $m$ by $r$, so that
\begin{equation}
n=3+q+r\ ,
\end{equation}
we have
\begin{eqnarray}
\Gamma _{\mu \nu mm}&=& \textstyle{2\over3}\,{\rm tr}\,(d_\mu \,d_\nu )
-\textstyle{\frac{1}{4}}r\,\delta _{\mu \nu}
=\textstyle{1\over3}\,d_{\rho mm}\,A_{\mu \rho ;\nu }\ ,\label{cGmuA}\\
\Gamma_{\mu \mu mn}&=& \textstyle{\frac{2}{3}} (d\,d)_{mn}
-\textstyle{\frac{1}{4}}(q+1)\,\delta_{mn}\ , \label{cGmut}\\
\Gamma _{ppmn}&=&\textstyle{\frac{2}{3}}(d\,d)_{mn}
-\textstyle{1\over 8}(r+2)\,\delta_{mn}
+\textstyle{\frac{1}{3}} d_{\mu pp}\,d_{\mu mn}\ , \label{cGmt}
\end{eqnarray}
where $(d\,d)_{mn}\equiv d_{\mu mp}\,d_{\mu np}$.

The remaining equations for which
the corresponding $\Gamma$ tensors vanish, $D_{m\mu\nu;\rho} =
D_{mnp;\mu} = D_{\mu mn;p}=0$, are solved by $A_{2m;\mu
}=A_{mn;p}=A_{\mu \nu ;p}=0$. Other solutions that
satisfy these equation, correspond to non-trivial invariances of
the $d_{abc}$ tensor.

Let us now turn again to the three cases discussed previously.
The only non-vanishing components of $d_{abc}$
are $d_{\mu mn}$ and
\begin{equation}
d_{222}= {1\over \sqrt2}\ , \quad
d_{2\mu\nu} = -{1\over\sqrt 2}\,
\delta_{\mu\nu}\ , \quad
d_{2mn}= {1\over 2\sqrt 2}\,\delta_{mn}\ , \label{parad}
\end{equation}
corresponding to
\begin{equation}
{\cal Y}(x) = {1\over \sqrt 2}\left(x_2^{\,3} -3 x_2\,(x_\mu^{\,2}
-\textstyle{1\over 2}\,x_m^{\,2} )\right) + 3d_{\mu mn}\,
x_\mu\,x_m\,x_n \ .                  \label{paray}
\end{equation}
The three cases are characterized by
the possible non-vanishing values of $d_{abb}$, which are
\begin{equation}
d_{2bb} = \frac{1}{2\sqrt 2} (r-2q), \quad \mbox{and}\quad d_{\mu bb}
= d_{\mu mm} .
\end{equation}

In case I we have $r=2q$, so that $n=3(q+1)$, and $d_{\mu mm}=0$. As
we established already, one must have
\begin{equation}
\Gamma_{\mu\nu mn} = \Gamma_{mnpq} = 0\ .
\end{equation}
Hence the $d_{\mu mn}$ may be regarded as $r\times r$ matrices, which
generate a $(q+1)$-dimensional Clifford algebra. In view of the
second condition, the dimension of this algebra is severely
constrained. According to \cite{BEC}, only $q= 1$, 2, 4 and 8 are
possible, corresponding to $n=6$, 9, 15 and 27, respectively.
This conclusion follows from the possible dimension of the
reducible representations of the Clifford algebra.
This case is related to Jordan algebras and the magic square
\cite{GuSiTo}. In addition we have the trivial case with $q=0$
and $n=3$.

For case II we have $d_{\mu mm}=0$ and $r-2q\not=0$.
It turns out that it is sufficient to restrict our analysis to
the case $q=-1$.
Then there are no indices $\mu$, so that
the non-vanishing coefficients $d_{abc}$ are
\begin{equation}
d_{222}={1\over\sqrt2}\ , \quad
d_{2mn}= {1\over2\sqrt2}\, \delta_{mn} \ ,
\end{equation}
with $r= n-2$ arbitrary. Obviously, we have
\begin{equation}
\Gamma_{mnpq} = - \textstyle{3\over8} \,\delta_{(mn}\,\delta_{pq)}\ ,
\quad H_{mn} = \delta_{mn}\ ,
\end{equation}
while all other components of $\Gamma$ vanish.

The reason why we do not have to consider $q\geq 0$, is that, after
identifying one of the indices $\mu$ with 3, we can always perform
a redefinition \eqn{ch23}. After this redefinition
we no longer have $d_{\mu mm}=0$, so that we can perform the
same steps as before, but now for case III. Nevertheless for
clarity of the presentation we briefly derive the consequences for
case II with arbitrary $q$.
We first use \eqn{GmH} to obtain
\begin{equation}
\Gamma^{(2)}_{mn} = \Gamma^{(2)}_{mp}\,H_{pn} ,\quad
\mbox{with} \quad
\Gamma^{(2)}_{mn} \equiv \Gamma_{mpqr}\,\Gamma_{pqrn} .
\label{Gam2H}
\end{equation}
Let us now decompose the space associated
with the indices $m, n, \ldots$
into the null space of $\Gamma^{(2)}$ and its orthogonal
complement. The
indices $m,n, \ldots$ are split accordingly
into indices $A, B, \ldots$ and $M, N,\ldots$, so that
$\Gamma^{(2)}_{Am} =\Gamma^{(2)}_{mA} =  0$ and
$\det \big(\Gamma^{(2)}_{MN}\big) \not= 0$.
This implies that
\begin{equation}
\Gamma_{mnpA} = 0,    \label{Gamvan}
\end{equation}
while \eqn{Gam2H} restricts the matrix $H$ according to
$H_{MA} = 0$ and $H_{MN} = \delta_{MN}$.
Combining $d_{\mu mm}=0$ and \eqn{cGmut},  \eqn{cGmt} and
\eqn{Gamvan}, we find
\begin{equation}
\Gamma_{\mu\mu AB} = \textstyle{1\over 8} (r-2q)\,\delta_{AB},
\qquad
\Gamma_{\mu\mu AM} =  \Gamma_{\mu\mu MA} =  0.
\end{equation}
{}From \eqn{GmuH} it then follows that the non-vanishing matrix
elements of $H$ are given by
$H_{AB} = - \delta_{AB}$ and $H_{MN}= \delta_{MN}$.
while
\begin{equation}
\Gamma_{\mu\nu MN} =   0\ . \label{Gresult}
\end{equation}
Therefore $\Gamma_{\mu\mu mn}$ is now fully known and non-vanishing.
On the other hand, $\Gamma_{\mu\mu mm}$ is
restricted to vanish by \eqn{cGmuA}. This implies that the
null space of $\Gamma^{(2)}$ is in fact empty, so that there are no
indices $A, B, \ldots$. Hence we find that $H_{mn} =
\delta_{mn}$. From \eqn{GmuH} it then follows that
$\Gamma_{\mu\nu mn}$ vanishes,
\begin{equation}
\Gamma_{\mu\nu m n}= 0\ ,
\end{equation}
while $\Gamma_{mnpq}$ remains arbitrary. Hence the coefficients
$d_{\mu mn}$ may again be regarded as $r\times r$ matrices generating
a $(q+1)$-dimensional Clifford algebra. This puts restrictions
on $r$ and $q$, but those are considerably weaker than
in the previous case.

Now let us turn to case III, where $d_{\mu mm}\not=0$.
By a suitable rotation of the components labeled by $\mu, \nu, \ldots$,
we choose the direction in which $d_{\mu mm}$ does not vanish to
be equal to $\mu=3$. The remaining  indices $\mu$ will be denoted
by $\hat\mu$. Subsequently we diagonalize $d_{3mn}$,
\begin{equation}
d_{3mn}=\sqrt{\textstyle\frac{3}{8}}\lambda _m\, \delta _{mn}\ .
\label{diag3}
\end{equation}
We then obtain from \eqn{cGmuA} that $A_{3\mu ;\nu }$ is symmetric in
$\mu$ and $\nu$ (this conclusion requires $d_{\mu mm}\not= 0$), which
implies that $A_{3\mu ;3}=0$. Substituting
this result into
\eqn{GmuA} for $\mu=\nu=3$, we obtain
\begin{equation}
\textstyle{1\over 4}(\lambda^2_m -1)
\delta_{mn}=    (\lambda _m-\lambda_n)A_{nm;3}\ .\label{33}
\end{equation}
As $A_{nm;3}$ is antisymmetric in $n$ and $m$, it follows
that both sides of the equation should vanish separately,
so that
\begin{equation}
\lambda_m^2 = 1\ , \qquad \Gamma_{\mu\nu mn}
= 0 \quad\mbox{for} \quad \mu=\nu=3\ .
\end{equation}
Splitting the range of indices $m,n,\ldots$ into indices
$x, y, \ldots$ and $\dot x, \dot y, \ldots$ such that
\begin{equation}
\lambda _x=1\ ;\qquad \lambda _{\dot x}=-1.
\end{equation}
it follows from \eqn{33} that $A_{x\dot y;3}=0$.
Subsequently, consider again
\eqn{GmuA} but now with $\mu=\hat\mu\not=3$,
$\nu =3$, $m=x$, $n=y$,
\begin{equation}
2d_{\hat\mu xy}=2d_{\hat \mu z(x}A_{y)z;3}+d_{\hat\rho xy}A_{\hat \mu
\hat\rho ;3}. \end{equation}
Multiplying the right-hand side with $d_{\hat\mu xy}$
gives zero by virtue of
the antisymmetry of the $A$ coefficients. This implies
$d_{\hat\mu xy}=0$. The same derivation
can be repeated for two dotted indices, so we are left with
the coefficients $d_{\hat\mu x\dot y}$ with mixed indices.
This then yields $\Gamma _{3\hat\mu mn}=0$.

In cases I and II we proved that
$\Gamma _{\mu \nu mn}=0$ in general. Therefore, in all cases with
$q\geq 0$, one can identify a suitable index $\mu =3$
and bring $d_{3mn}$ in diagonal form like in \eqn{diag3}, so that
one can employ the parametrization in terms of dotted and
undotted indices and
derive the restrictions for $d_{\hat\mu mn}$ as found above.
The present formulation is thus fully applicable to
all three cases with $q\geq0$ (for the moment we ignore
the results obtained above for the $A$ tensors, which apply only
to case III). Let us therefore proceed and present the relevant
equations in this formulation for the general case.

{}From $D_{3x\dot y;\hat\mu} = 0$ it follows that
\begin{equation}
A_{x\dot y;\hat\mu} = d_{\hat\nu x\dot y}\, H_{\hat\nu\hat\mu}\ ,
\end{equation}
where $H_{\hat\mu\hat\nu}\equiv \sqrt{2/3}\,
A_{3\hat\mu;\hat\nu}$.  In addition we have $ D_{\hat\mu x\dot y;3}=0$,
which can be solved by $A_{\dot x\dot y;3}=A_{xy;3}=A_{\hat\mu
\hat\nu;3}=0$ and has no consequences for the $d$ tensor.
When non-zero values are possible for the $A$ tensors, they
define non-trivial invariances of the $d$ coefficients.

Let us give the non-vanishing components of the
$\Gamma$ tensor in this notation  (cf. (\ref{defGmu}, \ref{defGm})),
\begin{eqnarray}
\Gamma _{\hat\mu \hat\nu xy}&=&\textstyle{\frac{2}{3}}\,
d_{\dot z x(\hat \mu}\,d_{\hat \nu )y\dot z}
-\textstyle{\frac{1}{4}}\,
\delta_{\hat\mu\hat\nu }\, \delta _{xy} \ ,\nonumber\\
\Gamma_{\hat\mu \hat\nu \dot x\dot y}&=&
\textstyle{\frac{2}{3}}\,d_{z \dot x(\hat \mu }\,
d_{\hat \nu )\dot y z}-\textstyle{\frac{1}{4}}\,
\delta_{\hat\mu \hat\nu }\,\delta_{\dot x\dot y} \ ,\nonumber\\
\Gamma_{x y\dot z\dot w} &=&
\textstyle{\frac{2}{3}}\,d_{\hat\mu x(\dot z} \,
d_{\dot w)y\hat\mu }-\textstyle{\frac{1}{4}} \,
\delta _{xy} \,\delta _{\dot z\dot w} \ .\label{defGamhat}
\end{eqnarray}
We denote the range of indices $\hat\mu$, $x$ and $\dot x$ by $q$,
$p$ and $\dot p$, respectively, so that $r=p+\dot p$ and $n= 3+
q+ p+\dot p$.

These equations have a remarkable symmetry under interchange of the
indices $\hat \mu $, $x$ and $\dot x$. This is not a coincidence
and is related to the redefinitions that were explained previously.
To see this, consider the cubic polynomial $\cal Y$, which has
acquired the following form (for $q\geq 0$),
\begin{eqnarray}
{\cal Y}(x) &=&\frac{1}{\sqrt{2}}\,x_2\,(x_2 +\sqrt3\, x_3)\,
(x_2 -\sqrt 3\, x_3)\nonumber\\
&&+ \frac{3}{\sqrt{2}}\left( -x_2\,x_{\hat \mu }^2+\half
(x_2+\sqrt{3}\,x_3)\,x_x^2
+\half (x_2-\sqrt{3}\,x_3)\,x_{\dot x}^2 \right)\nonumber\\
&& +6\,d_{\hat \mu m\dot m}\,x_{\hat \mu}\,x_x\,x_{\dot x}.
\label{paray2}
\end{eqnarray}
The replacement \eqn{ch23} induces an interchange of the quantities
$x_2$ and $-\half(x_2 \pm \sqrt 3 x_3)$, which leaves the form of
the function ${\cal Y}(x)$ unchanged, except that the labels
$\hat\mu$, $x$ and $\dot x$ are interchanged. Similarly, the
replacement \eqn{ch3m} corresponds to an interchange of labels
$x$ with $\dot x$ (of course, the range of the various indices
changes accordingly).

Case I is now characterized by $q=p=\dot p$, case II
by $p=\dot p\not= q$, and case III by $p\not=\dot p$. Contraction of
the above tensors leads to the following equations,
\begin{eqnarray}
&&\Gamma_{\hat\mu\hat\mu xy} = \Gamma_{\dot z\dot z xy} +
\textstyle{1\over4}(\dot p-q)\,\delta_{xy}\ , \nonumber\\
&&\Gamma_{\hat\mu\hat\mu \dot x\dot y} = \Gamma_{z z\dot x\dot y} +
\textstyle{1\over4}(p-q)\,\delta_{\dot x\dot y}\ , \nonumber\\
&&\Gamma_{\hat\mu\hat\nu xx} = \Gamma_{\hat\mu\hat\nu \dot x\dot x} +
\textstyle{1\over4}(\dot p-p)\,\delta_{\hat\mu\hat\nu}\ .
\label{Gamcontr}
\end{eqnarray}
In this notation, the equations (\ref{GmuH}-\ref{GmH}) take the form
\begin{eqnarray}
&&\Gamma_{\hat \mu\hat \nu x y} = \Gamma_{\hat \rho\hat \nu
xy}\,H_{\hat \rho\hat \mu} = -\Gamma_{\hat \mu\hat \nu zy}\, H_{zx}\ ,
\nonumber \\ &&\Gamma_{\hat \mu\hat \nu \dot x\dot y} =
-\Gamma_{\hat \rho\hat \nu \dot x\dot y}\, H_{\hat \rho\hat \mu}
= -\Gamma_{\hat \mu\hat \nu \dot z\dot y}\, H_{\dot z\dot x}\ , \nonumber
\\ &&\Gamma_{x y\dot x\dot y}= \Gamma_{zy\dot x\dot y}
\, H_{zx} = \Gamma_{xy\dot z\dot y }
\, H_{\dot z\dot x}\  ,  \label{GHHH}
\end{eqnarray}
where we suppressed the equations involving $H_{x\dot y}$ and
$H_{\dot x y}$, which have no consequences for the
$d$-coefficients.

The symmetry noted above should be taken into account when
identifying inequivalent $d$ tensors. However, its presence also
facilitates our work, as it allows us to apply the following
lemma in three possible situations: \\

\noindent
{\bf Lemma}: {\it
 Consider one of the three matrices $H$, say, $H_{\hat
\mu\hat\nu}$. Then, either
the two other matrices $H$ are of equal dimension
($p=\dot p$), in which case $\Gamma_{\hat\mu\hat\nu xx} =
\Gamma_{\hat\mu\hat\nu \dot x\dot x}=0$, or they
are not of equal dimension ($p\not=\dot p$), in which case
$H_{\hat\mu\hat\nu}$ is equal to plus or minus
the identity matrix, with
$\Gamma_{\hat\mu\hat\nu xy}$ or
$\Gamma_{\hat\mu\hat\nu \dot x\dot y}$ vanishing, respectively.}\\

\noindent
To prove this lemma, multiply the third equation (\ref{Gamcontr})
with $H_{\hat\mu\hat\rho}$ and apply (\ref{GHHH}).
When $p=\dot p$
the corresponding equations lead to  $\Gamma_{\mu\nu xx} =
\Gamma_{\mu\nu \dot x\dot x}=0$, as claimed above. On the other
hand, when $p\not=\dot p$,
it follows that $H_{\hat\mu\hat\nu}$ is a symmetric matrix which
can be diagonalized. Consider first
the case where $\hat\mu$ and $\hat\nu$
belong to an eigenspace of $H$ with
eigenvalue different from $\pm 1$. Then it follows from (\ref{GHHH})
that $\Gamma_{\hat\mu\hat\nu xx} =
\Gamma_{\hat\mu\hat\nu \dot x\dot x} =0$, which leads to $p=\dot p$
and thus to a contradiction. Hence $H_{\hat\mu\hat\nu}$ has only
eigenvalues equal to $\pm 1$. Assume now that both eigenvalues occur.
Consider then indices $\hat\mu$ and $\hat\nu$ corresponding to
the subspace with
eigenvalue $+1$, and an index $\hat\rho$ belonging to the subspace
with eigenvalue $-1$. Then (\ref{GHHH}) implies that (no sum over repeated
$\hat \rho $ index) \begin{equation}
\Gamma_{\hat\mu\hat\nu \dot x\dot y} =
\Gamma_{\hat\rho\hat\rho x y} =
\Gamma_{\hat\mu\hat\rho \dot x\dot y} =
\Gamma_{\hat\mu\hat\rho x y} =
\Gamma_{\hat\nu\hat\rho \dot x\dot y} =
\Gamma_{\hat\nu\hat\rho x y} =  0 \ .
\end{equation}
According to the last four equations
$d_{\hat\rho}$ anticommutes as
a matrix with $d_{\hat\mu}$ and $d_{\hat\nu}$. Thus we perform
the following calculation (no sum over repeated
$\hat \rho $ index),
\begin{eqnarray}
0&=&d_{\hat\rho x \dot x}\,\Gamma _{\hat \mu \hat\nu\dot x\dot y}\,
d_{\hat\rho\dot yy} \nonumber\\
&=&\textstyle{\frac{2}{3}}\big( d_{\hat\rho}\, d_{(\hat \mu }
\, d_{\hat \nu )}\, d_{\hat\rho} \big)_{xy}
-\textstyle{\frac{1}{4}}\left(d_{\hat\rho}\,d_{\hat\rho}\right)
_{x y}\,\delta _{\hat \mu \hat \nu }\nonumber\\
&=&\Gamma _{\hat \mu \hat \nu
x z}\, (d_{\hat\rho}\,d_{\hat\rho})_{z y}
=\textstyle{\frac{3}{8}}\,\Gamma _{\hat \mu
\hat \nu x y} \ .
\end{eqnarray}
Hence both $\Gamma_{\hat\mu\hat\nu xy}$ and
$\Gamma_{\hat\mu\hat\nu \dot x\dot y}$ vanish, which requires that
$p=\dot p$, thus leading to a contradiction. Hence the eigenvalues
of $H$ must all be equal, which completes the proof of the lemma.\QED

With the help of this lemma it is straightforward to analyze the
various solutions of (\ref{defGamhat},\ref{GHHH}). First we
assume that $q$, $p$ and $\dot p$ are non-vanishing. Application of the
lemma then reveals that there are no solutions with different
values for $q$, $p$ and $\dot p$, simply because two $\Gamma$
tensors must then vanish, which, by (\ref{Gamcontr}) implies that
at least two of the parameters $q$, $p$ or $\dot p$ should be equal.
Because of the symmetry we can choose either two of the
parameters equal. Let us assume, for instance, \underline{$q\neq
p= \dot p\neq 0$}. Then, from the lemma applied to the three
matrices $H$ one finds four possibilities, two of which implying
that two uncontracted $\Gamma$ tensors vanish, which is
inconsistent with $q\not=p$ or $q\not=\dot p$. Then one has the
third possibility corresponding to
\begin{equation}
p=\dot p\ :\quad
\Gamma _{xy\dot x\dot y}=\Gamma _{\hat\mu\hat\nu xx}
=\Gamma _{\hat\mu\hat\nu \dot x\dot x}=0\ , \quad
H_{xy} =- \delta_{xy} \ , \quad H_{\dot x\dot y} = -\delta_{\dot
x\dot y}\ .
\end{equation}
On the other hand the first line of \eqn{Gamcontr} implies that
$\Gamma _{\hat \mu \hat \mu xx}=\textstyle{\frac{1}{4}}(p-q)p$,
which must vanish according to the above equation. Hence $p=0$;
this is one of the cases to be discussed below. The remaining
possibility, which leaves $q$ arbitrary corresponds to
\begin{equation}
p=\dot p\ :\quad \Gamma_{\hat\mu\hat\nu xy} =
\Gamma_{\hat\mu\hat\nu \dot x\dot y} =  0\ ,\quad
H_{xy} = \delta_{xy} \ , \quad H_{\dot x\dot y} =
\delta_{\dot x\dot y}\ .  \label{resp=dp}
\end{equation}
Clearly, this solution belongs to case II, while the equivalent
solution with $q=p\neq \dot p$ or $q=\dot p\neq p$ belongs to
case III.

The case \underline{$q=p=\dot p$} is case I, for which we showed already
before that all $\Gamma $ symbols are zero with traceless $d$ coefficients.

What remains is to investigate the situation where
at least one of the parameters $q$, $p$ or $\dot p$ vanishes (this may
occur in case I, II or III depending on the values of the other
two parameters).  In that case only one of the tensors $\Gamma$
remains (unless one of the other parameters vanishes as well).
Let us choose \underline{$q=0$}. There is only
\begin{equation}
q=0\ :\quad \Gamma_{xy\dot x\dot y} = -\textstyle{1\over 4}
\,\delta_{xy}\,\delta_{\dot x\dot y}\ , \quad H_{xy} =
\delta_{xy}\ ,\quad H_{\dot x\dot y} = \delta_{\dot x \dot y}\ ,
\end{equation}

This completes the classification of the coefficients $d_{abc}$ satisfying
the equation (\ref{GamD}).
\newpage

\section{Results of the classification.}
\setcounter{equation}{0}
\subsection{$d$-coefficients and Clifford algebras.}
In the previous section we obtained the possible tensors
$d_{abc}$ that are solutions to \eqn{GamD}, up to arbitrary
$O(n\!-\!1)$
rotations. The indices $a,b,\ldots$, are decomposed into indices
2, $\mu $ and $m$, where $\mu $ and  $m$ take $q+1$ and
$r$ values, respectively. We thus have $n=3+q+r$.

The general results for the $d$ tensors are summarized in
\eqn{parad} and (\ref{paray}), where, as we shall see shortly,
the coefficients $d_{\mu mn}$ satisfy the defining
relation (up to a proportionality factor) of the generators of
a Clifford
algebra and can thus be expressed as (symmetric, real) gamma matrices
according to
\begin{equation}
d_{\mu mn}=\sqrt{\textstyle\frac{3}{8}} \left( \gamma _\mu \right)
_{mn}.\label{gamma}
\end{equation}
Therefore the function ${\cal Y}(x)$ acquires the generic form
\begin{equation}
{\cal Y}(x) = {1\over \sqrt 2}\left\{x_2^{\,3} -3 x_2\,(x_\mu^{\,2}
-\textstyle{1\over 2}\,x_m^{\,2} ) + \textstyle{3\over
2}\sqrt{3}\left(\gamma _\mu\right) _{mn}\, x_\mu\,x_m\,x_n
\right\}\ ,
\label{parayf} \end{equation}
where the gamma matrices generate a real representation of the Clifford
algebra ${\cal C}(q+1,0)$ with positive metric. Let us now
analyze the various solutions found in the previous section.

The first case is \underline{$q=-1$} (i.e., indices  $\mu$ are
absent). As  $d_{\mu mn}$ does not exist, the $d$ coefficients
are completely given by \eqn{parad}, and the corresponding
function $\cal Y$ reads
\begin{equation}
{\cal Y}(x) = {1\over \sqrt 2}\left(x_2^{\,3} +\textstyle{3\over
2}x_2\,x_m^{\,2} \right) \ .
\end{equation}
As shown previously, $\Gamma_{mnpq}= -{3\over 8}\delta_{(mn}\,
\delta_{pq)}$, which vanishes only for $r=0$. We denote
these solutions by $L(-1,r)$ with $r\geq 0$ and $n=2+r$.

For $q\geq 0$ there is one value of $\mu $ which we
denote by "3" and we split indices $m$ into $x$ or $\dot x$,
taking $p$ and $\dot p$ values, respectively. These indices are
distinguished by
\begin{equation}
d_{3xy}=\sqrt{\textstyle\frac{3}{8}}\, \delta _{xy}\ ,\qquad
d_{3\dot x\dot y}=-\sqrt{\textstyle\frac{3}{8}}\, \delta _{\dot
x\dot y}\ , \qquad  d_{3x\dot y}=0.
\end{equation}

For \underline{$q=0$} there is no further restriction; $p$ and
$\dot p$ are arbitrary
and we denote the solution as $L(0,P,\dot P)\equiv L(0,\dot P,
P)$, where we replace $p$ and $\dot p$ by $P$ and $\dot P$ in
order to have a uniform
notation for the reducible representations of the Clifford
algebra (to be discussed below). Whenever $P$
or $\dot P$ are zero we write $L(0,P)\equiv L(0,P,0)$. The
diagonal matrix
\begin{equation}
\left( \gamma_3\right) _{mn}\equiv \sqrt{\textstyle\frac{8}{3}}\,
d_{3mn} \ ,
\end{equation}
can be viewed as a gamma matrix that generates a
one-dimensional
Clifford algebra ${\cal C}(1,0)$. This algebra has two inequivalent
irreducible representations corresponding to $+1$ and $-1$. The
numbers $P$ and $\dot P$ specify the multiplicities of these
representations in $\gamma_3$. The corresponding function $\cal
Y$ follows directly from \eqn{parayf},
\begin{eqnarray}
{\cal Y}(x) &=&\frac{1}{\sqrt{2}}\Big\{x_2\,(x_2 +\sqrt3\, x_3)\,
(x_2 -\sqrt 3\, x_3)\nonumber\\
&&\qquad + {\textstyle{3\over 2}} (x_2+\sqrt{3}\,x_3)\,x_x^2
+{\textstyle{3\over 2}} (x_2-\sqrt{3}\,x_3)\,x_{\dot x}^2
\Big\}\ .
\end{eqnarray}
The non-vanishing $\Gamma$ tensor is
\begin{equation}
\Gamma_{xy\dot x\dot y}= -{\textstyle{1\over 4}} \delta_{xy}\,
\delta_{\dot x\dot y}\ ,
\end{equation}
which vanishes whenever $p$ or $\dot p$ vanishes. Note that we
have $n=3+p+\dot p$.

For \underline{$q>0$} we may restrict ourselves to $r>0$, as the case $q>0,\
r=0$ is equivalent to $L(0,q,0)$ by a rotation \eqn{ch23}.
Denoting the values of $\mu\not=3$ by $\hat \mu$, the tensors
$d_{\hat\mu mn}$ satisfy
\eqn{resp=dp}, which implies that we can define the following
$r\times r$  gamma matrices,
\begin{equation}
\gamma _{\hat \mu } = \sqrt{\textstyle\frac{8}{3}}
\pmatrix {0& d_{\hat \mu x\dot v}\cr
d_{\hat\mu \dot y w}&0\cr } \ ,
\qquad
\gamma _{3} = \sqrt{\textstyle\frac{8}{3}}\pmatrix {1&0 \cr
0&-1\cr}\ .
\end{equation}
We have thus established (\ref{parayf}). To classify all cases
with $q>0$ one must consider all possible gamma matrices that
generate a real Clifford algebra ${\cal C}(q+1,0)$. The
irreducible representations (with positive definite metric) are
listed in table~\ref{Cliffp0} \cite{CliffordR}.
\begin{table}[tb]
\begin{center}
\begin{tabular}{||c|c|c|c|c|c||}\hline
$q$ &$q+1$ &${\cal C}(q+1,0)$& ${\cal D}_{q+1}$&$\Rbar ({\cal D}_{q+1})$&
${\bf C}$ \\ \hline &&&&&\\[-3mm]
$-1$ & 0&$\Rbar$    &1         &$\Rbar $ &$\Rbar$     \\
0    & 1&$\Rbar\oplus \Rbar $&1&$\Rbar $&$\Rbar$ \\
1    & 2&$\Rbar(2)$ &2         &$\Rbar (2)$&$\Rbar$     \\
2    & 3&$\Cbar(2)$ &4         &${\cal C}(3,1)$& $\Cbar$     \\
3    & 4&$\Hbar(2)$ &8         &$\Hbar \otimes \Hbar (2)$& $\Hbar$
\\ 4    & 5&$\Hbar (2)\oplus \Hbar (2)$&8
        &$\Hbar \otimes \Hbar (2)$&  $\Hbar$\\
5    & 6&$\Hbar(4)$ &16&$\Hbar \otimes \Hbar (4)$&  $\Hbar$   \\
6    & 7&$\Cbar(8)$ &16&${\cal C}(8,0)$        & $\Cbar$     \\
7    & 8&$\Rbar(16)$&16&$\Rbar (16)$        & $\Rbar$    \\
$n+7$ & $n+8$ & $\Rbar(16)\otimes{\cal C}(n,0)$&16 ${\cal D}_n$ & $\Rbar
(16)\otimes \Rbar ({\cal D}_n)$& as for $n$\\[1mm]
\hline
\end{tabular}
\end{center}
\caption{Real Clifford algebras ${\cal C}(q\!+\!1,0)$. Here
${\bf F}(n)$ stands for $n\times n$
matrices with entries over the field
$\bf F$, while ${\cal D}_{q+1}$ denotes the
real dimension of an irreducible representation of
the Clifford algebra. We decompose the matrices $\Rbar ({\cal
D}_{q+1})$ acting on the real irreducible
representation space, either as a direct product
with the Clifford algebra {\em representation} as a factor, or in
the form of a higher-dimensional Clifford algebra. This
decomposition is used to determine the centralizer $\bf C$ of the
Clifford algebra in this representation.}
\label{Cliffp0}
\end{table}
They are unique
except when the Clifford module consists of a direct sum of two
factors. As shown in table~\ref{Cliffp0} this is the case for
$q=0$ mod 4, where there exist two inequivalent irreducible
representations\footnote{
These are the only dimensions for which the product of all
gamma matrices, $Q\equiv \gamma_1\cdots\gamma_{q+1}$, commutes
with every individual matrix and has square $\unity$; the two
inequivalent representations are related by an overall sign
change in the gamma matrices: $\gamma_\mu\to - \gamma_\mu$, so
that $Q$ changes from $+\unity $ to $-\unity $, or vice
versa.}. This implies
that for $q\neq 0$ mod 4, the gamma matrices are unique once we
specify the number of irreducible representations. The solution
for the $d$ coefficients is then denoted by $L(q,P)$, where $P$
denotes the number of irreducible representations. We have thus
$r=P\, {\cal D}_{q+1}$, or, equivalently, $n=3+q+P\,{\cal
D}_{q+1}$.
However, when $q$ is a multiple of 4 (i.e., $q=4m$ with $m$ integer),
then there exist two inequivalent irreducible representations and
the solutions are characterized by specifying the multiplicities
$P$ and $\dot P$ of each of the two representations.  The
solutions are therefore denoted by $L(4m,P,\dot P)\equiv L(4m, \dot P,
P)$ and we have $n=3+4m+(P+\dot P){\cal D}_{4m+1}$. Whenever $P$
or $\dot P$ vanishes, we denote the solutions by $L(4m,P)\equiv
L(4m,P,0)$.

This concludes the classification of the various solutions.
The only components of $\Gamma _{abcd}$ that possibly differ from zero
are
\begin{equation}
\Gamma _{mnpq}={\textstyle\frac{3}{8}}\left[ \left( \gamma _\mu
\right) _{(mn} \left( \gamma _\mu \right) _{pq)} -
\delta_{(mn}\delta _{pq)}\right] .
\end{equation}
As one easily verifies, this tensor vanishes only for $L(-1,0)$,
$L(0,r)$, $L(1,1)$, $L(2,1)$, $L(4,1)$ and $L(8,1)$,
corresponding to $n= 2$, $3+r$, 6, 9, 15 and 27, respectively.
Note that the contracted tensor
\begin{eqnarray}
\Gamma_{mnpp}&=&\textstyle{\frac{1}{8}} (2q-r)\, \delta_{mn}\qquad
\mbox{for }q\neq 0 \\
&=& -\textstyle{\frac{1}{8}}P\left(\delta
_{mn} -\left(\gamma_3\right)
_{mn}\right) -\textstyle{\frac{1}{8}}\dot P\left(\delta
_{mn} +\left(\gamma_3\right) _{mn}\right)\qquad
\mbox{for }q=0 \nonumber
\end{eqnarray}
($q=0$ is special, because it represents the only case where a gamma
matrix can have a non-zero trace) has only
zero eigenvalues in those cases where we already know that
$\Gamma _{mnpq}=0$. This
implies that the equation $\Gamma _{mnpq} Z^q=0$ has only
non-trivial solutions $Z^q$ when $\Gamma_{mnpq}$ vanishes.

\subsection{$A$-coefficients and symmetry groups}
Now that we have found the non-vanishing components of $\Gamma
_{abcd}$ we consider the solutions for the corresponding tensors
$A_{ab;c}$ as defined by \eqn{GamD}. They are determined modulo
solutions of the homogeneous equation,
\begin{equation}
d_{d(ab}\,A_{c)d} = 0, \label{homeq}
\end{equation}
which define the invariances of the coefficients $d_{abc}$.
We recall that these symmetries must
preserve the metric $\delta _{ab}$, so that
the matrices $A_{ab}$ are antisymmetric.
There are only two types of invariances. First there is
\begin{equation}
A_{2m}=-A_{m2} =\sqrt{3}\,\zeta_m\ , \qquad A_{m\mu}= -A_{\mu m} =
\left( \gamma _\mu\right) _{mn}\, \zeta_n \ ,\label{solzetA}
\end{equation}
where $\zeta_m$ must satisfy
\begin{equation}
\Gamma_{mnpq}\,\zeta_q = 0\ .
\end{equation}
However, from the discussion at the end of the previous subsection it
follows that this equation has only non-trivial solutions for $\zeta _m$
when $\Gamma_{mnpq}$ vanishes.

The second type of solutions of (\ref{homeq}) corresponds to
the invariance group of the tensor $d_{\mu mn}\propto \gamma_{\mu mn}$
associated with the matrices $A_{\mu\nu}$ and $A_{mn}$,
\begin{equation}
A_{\mu\nu}\, \left( \gamma_\nu\right) _{mn} +
\gamma_{\mu p(m}\, A_{n)p} =0 .
\label{HOq+1}
\end{equation}
For any $A_{\mu \nu }$ there is the solution
\begin{equation}
A_{mn} = {\textstyle{1\over4}} A_{\mu\nu} \,
\big(\gamma_\mu\gamma_\nu\big)_{mn} .
\label{cover}
\end{equation}
Obviously, the group
associated with $A_{\mu\nu}$ is the rotation group $SO(q\!+\!1)$,
which acts on the spinor coordinates labeled by $m$ according to
its cover group. Besides there can be additional invariances
that act exclusively in spinor space and commute with the gamma
matrices and thus with the corresponding representation of the
Clifford algebra. Hence we are interested in the metric-preserving
elements of the centralizer of the Clifford algebra in the
$r$-dimensional real representation (i.e., the
antisymmetric matrices $A_{mn}$ belonging to $\Rbar(r)$ that
commute with $\gamma_\mu$). Let us first determine the
centralizers for the irreducible representations.

According to
Schur's lemma, matrices that commute with an {\em irreducible}
representation of the Clifford algebra must form a division
algebra. Table~\ref{Cliffp0} lists the centralizers of the
real irreducible representations, which are thus equal to
$\Rbar$, $\Cbar$ or $\Hbar$. We briefly present the arguments
leading to this result\footnote{Many results
on real irreducible representations of the Clifford algebras and their
centralizers have been explicitly worked out in \cite{Okubo}.  Another
useful reference is \cite{Coq}.}.
First consider $q+1$ even.  The only commuting element in the Clifford
algebra representation is $\unity $, while the centralizer is just the
factor in $\Rbar({\cal D}_{q+1})$ that
multiplies the Clifford algebra representation
(cf. table~\ref{Cliffp0}).  In this way we find that the centralizer
is $\Rbar$ for $q+1=0$ or 2 mod 8, and $\Hbar$ for $q+1=4$ or 6
mod 8. Now take $q+1$ odd. For $q=4m$ the irreducible
representation of the Clifford algebra corresponds to only one of
the terms of the direct sum in ${\cal C}(4m+1)$.
Just as above, the only commuting element in this
representation is $\unity$, and the
centralizer is obtained as the factor that multiplies the
Clifford algebra representation in
$\Rbar ({\cal D}_{q+1})$, i.e.,  $\Rbar $
for $q+1=1$ mod 8 and $\Hbar $ for $q+1=5$ mod 8.
What remains is the case $q=2+4m$. Then, as indicated in
table~\ref{Cliffp0}, the representation space is isomorphic to a
higher-dimensional Clifford algebra,
which makes it easy to verify that only
$\unity $ and $Q\equiv \gamma _1\cdots\gamma _{q+1}$
span the centralizer.
(Note that for $q=2+4m$, $Q^2=-\unity $, while for $q=4m$, $Q$ itself is
represented by $\pm \unity $). We thus conclude that the centralizer is
equal to $\Cbar$ for $q+1=3$ or 7 mod 8.

To analyze the reducible representations, we first rewrite $\Rbar
(r)$ as $\Rbar(p)\otimes \Rbar ({\cal D}_{q+1})$, where $p$ is
the number of irreducible representations, thus $p=P$ for
$q\neq 4m$ and $p=P+\dot P$ for $q=4m$.
Consider first $q\neq 4m$ such
that $\gamma _\mu =\unity _P\otimes \gamma_\mu ^{irr}$.
This shows that the centralizer is the direct product of $\Rbar
(P)$ with the centralizer of $\gamma_\mu ^{irr}$. leading to
$\Rbar (P)$ for
$q=1,7$ mod 8, to $\Cbar (P)$ for $q=2,6$ mod 8, and to $\Hbar
(P)$ for $q=3,5$ mod 8.  What remains are the cases $q=0$ mod 4,
when we have $\gamma _\mu
=\eta \otimes \gamma _\mu ^{irr}$, where $\eta =\mbox{diag}(1,
\ldots 1,-1,\ldots - 1)$.  Writing $A_{mn}$ as $A=H\otimes S$,
where $H\subset \Rbar(p)$ and $S\subset\Rbar({\cal D}_{q+1} )$,
we have the condition
\begin{equation}
[A,\gamma _\mu ]= H\eta \otimes S\gamma _\mu -\eta H\otimes \gamma _\mu
S=0. \end{equation}
In the sector proportional to
$(\unity \pm\eta)H(\unity\mp\eta)$,
it follows that $S$ anticommutes with
$\gamma_\mu ^{irr}$; $S\,S^T$ is then a symmetric matrix
that commutes with
$\gamma_\mu ^{irr}$, so that it must be proportional to $\unity$.
Therefore $S$ is orthogonal and $S\,\gamma_\mu ^{irr}S^{-1}
=-\gamma_\mu ^{irr}$. This leads to a contradiction, as it
implies that $\gamma_\mu ^{irr}$ and $-\gamma_\mu ^{irr}$ are
equivalent representations. Consequently the matrices $H$ are
restricted to the $\Rbar (P)\oplus\Rbar (\dot P)$ matrices
commuting with $\eta $. For these matrices the same
considerations apply as for $q\neq 4m$. The result is then that
the centralizer is the direct product of $\Rbar (P)\oplus
\Rbar (\dot P)$ with the
centralizer of $\gamma_\mu ^{irr}$, which corresponds to
$\Rbar (P)\oplus \Rbar (\dot P)$ for $q=0$ mod 8, and $\Hbar
(P)\oplus \Hbar (\dot P)$ for $q=4$ mod 8.

Now we determine the antisymmetric matrices in these centralizers
corresponding to the generators of the metric-preserving
subgroups. In each case these
centralizers can be written as the direct product of real
matrices with a division algebra (in the real representation, so
that the imaginary units become antisymmetric matrices).
Therefore in
the complex or the quaternionic representation the antisymmetry
requirement takes the form of an antihermiticity requirement. The
metric-preserving groups are therefore
\begin{eqnarray}
\mbox{for }q=1,7 \mbox{ mod }8&:& SO(P)\nonumber\\
\mbox{for }q=0 \mbox{ mod }8&:& SO(P)\otimes SO(\dot P)\nonumber\\
\mbox{for }q=2,6 \mbox{ mod }8&:& U(P)\nonumber\\
\mbox{for }q=3,5 \mbox{ mod }8&:& U(P,\Hbar )\equiv U\!Sp(2P)\nonumber\\
\mbox{for }q=4 \mbox{ mod }8&:& U\!Sp(2P)\otimes U\!Sp(2\dot P)
\label{mpgc}
\end{eqnarray}
In conclusion, we summarize the symmetries of the tensors
$d_{abc}$. First there are the symmetries \eqn{solzetA} for the
cases $L(-1,0)$, $L(0,r)$, $L(1,1)$, $L(2,1)$, $L(4,1)$ and $L(8,1)$.
Secondly there is the group $SO(q+1)$ and the group mentioned in
\eqn{mpgc} represented by matrices $S_{mn}$. This gives
\begin{eqnarray}
A_{\mu \nu }&=& \mbox{arbitrary}\nonumber\\
A_{mn}&=& \textstyle{\frac{1}{4}}\left( \gamma_\mu \gamma _\nu \right)
_{mn}A_{\mu \nu }+S_{mn}\nonumber\\
A_{2m}&=&\sqrt{3}\,\zeta _{m}\nonumber\\
A_{m\mu }&=&\left( \gamma _\mu \right) _{mn}\zeta_n  .
\label{gensolA}
\end{eqnarray}

Now that we have determined the solutions of the homogeneous equation
\eqn{homeq}, we turn to the inhomogeneous equation \eqn{GamD}. A
particular solution is
\begin{equation}
A_{2m;n}=-A_{m2;n} ={\textstyle{3\over 4}} \sqrt 2\, \delta_{mn}\ ,
\qquad  A_{m\mu;n}= -A_{\mu m;n} = \textstyle{\frac{1}{4}}\sqrt{6}
\left( \gamma _\mu\right) _{mn} \ .\label{specsolA} \end{equation}
When $\Gamma_{mnpq}= 0$ these solutions correspond to an
invariance of the $d_{abc}$ coefficients and are already contained
in the previous transformations.

\section{Implications for homogeneous special spaces}
\label{conclusions}
\setcounter{equation}{0}
Now we return to special geometry and the cubic
polynomial $C(h)$, defined in (\ref{Cpoly}). Using the canonical
parametrization, we first introduce an extra coordinate $x_1$,
and add the corresponding terms $x_1^{\,3} - {1\over 2}x_1\,
x_a^{\,2}$ to the polynomial (\ref{cubic}). Giving up the canonical
parametrization, we no longer have to restrict ourselves to $O(n-1)$
redefinitions, and we can make arbitrary
linear redefinitions of the $x_1,\ldots,x_n$.  Using
\begin{eqnarray}
&&h^1 = 3^{-1/3}\,\big(x_1+\sqrt 2 \,x_2\big)\ ,  \nonumber\\
&&h^2 = 3^{-1/3}\,\big(x_1-{\textstyle{1\over 2}}\sqrt 2\,
x_2\big)\ ,  \nonumber \\
&&h^\mu =2^{-1/2}\cdot 3^{1/6}\, x_\mu \ , \qquad h^m =2^{-1/2}\cdot
3^{1/6}\, x_m \ , \end{eqnarray}
the polynomial $C(h)$ acquires the generic form given in section
1 (cf. \ref{genC1}),
\begin{equation}
C(h) = 3\Big\{ h^1\,
\big(h^2\big)^2 -h^1\,\big(h^\mu\big)^2 -h^2\,\big(h^m\big)^2
+\gamma_{\mu mn}\,h^\mu\, h^m\,h^n\Big\} \ . \label{genC}
\end{equation}
We stress that this parametrization no longer coincides
with the canonical one. The possible realizations for the gamma
matrices were discussed in the previous section. Note that we have
\begin{equation}
n= 3+q+r\ ,\qquad \mbox{with}\quad r=(P+\dot P)\,{\cal D}_{q+1}\ ,
\end{equation}
where the integers $P$ and $\dot P$ characterize the
representations for the gamma matrices, as discussed in the
previous section.

We now summarize the linear transformations of $h^A$ that leave
(\ref{genC}) invariant. They can either be determined directly from
\eqn{genC}, or can be evaluated from $\delta x^A = B^A{}_B x^B$,
using \eqn{cBpar} with $A_{ab}=B^c{}_1 \,A_{ab;c}\,$, where
$A_{ab;c}$ is taken from \eqn{specsolA}, plus a homogeneous
solution as in \eqn{gensolA},
\begin{eqnarray}
\delta h^1 &=& 2\xi_2\,h^1 + 2\xi_m h^m   \ , \nonumber \\
\delta h^2 &=& -\xi_2\,h^2 -\zeta_m\, h^m+ 2\xi_\mu\,h^\mu \ ,
\nonumber \\
\delta h^\mu &=& -\xi_2\,h^\mu + 2\xi_\mu \,h^2 -
\zeta_n\,\gamma_{\mu mn}\,h^m + A_{\mu\nu}\,h^\nu \ , \label{htrans} \\
\delta h^m &=& {\textstyle{1\over 2}}\xi_2\,h^m + \xi_m \,h^2 -
\zeta_m\,h^1 + \xi_n \,\gamma_{\mu mn}\,h^\mu + \xi_\mu\,
\gamma_{\mu mn}\,h^n + A_{mn}\,h^n\ . \nonumber
\end{eqnarray}
The symmetries
corresponding to the parameters $\zeta_m$ only exist
when the tensor $\Gamma_{mnpq}$ vanishes.  As before, $A_{mn}$ and
$A_{\mu\nu}$ are antisymmetric matrices that leave the gamma matrices
invariant (cf.  (\ref{HOq+1})).  As explained in the previous section,
$A_{\mu\nu}$ and $A_{mn}$ generate the product of $SO(q\!+\!1)$ and the
metric-preserving group in the centralizer of the corresponding Clifford
algebra representation given in \eqn{mpgc}.  The parameters are
defined as
follows \begin{eqnarray}
&& B^1_2= \sqrt 2\,\xi_2 \ ,\qquad
B^1_m = \sqrt{\textstyle{2\over 3}}\,
\big(\xi_m-\zeta_m\big)\ ,\qquad
B^1_\mu = 2 \sqrt{\textstyle{2\over 3}}\, \xi_\mu \ , \\
&& A_{2m} =-A_{m2}= {\textstyle{1\over 2}}\sqrt 3\,\big(\xi_m +
\zeta_m\big)\ , \qquad  A_{m\mu} =-A_{\mu m}={\textstyle{1\over 2}}
\gamma_{\mu mn}\, \big(\xi_m + \zeta_m\big)\ . \nonumber
\end{eqnarray}

It is illuminating to decompose the generators with respect to
the abelian generator $e_2$ associated with the parameter
$\xi_2$. The algebra then decomposes according to
\begin{equation}
{\cal X} = {\cal X}_{-3/2} + {\cal X}_0 + {\cal X}_{3/2} \ ,
\end{equation}
where ${\cal X}_{-3/2}$ contains the generators associated with
the parameters $\zeta_m$ (which is thus only present when
$\Gamma_{mnpq}=0$), ${\cal X}_0$ consists of the generators
associated with $\xi_2$, $\xi_\mu$, $A_{\mu\nu}$ and $A_{mn}$, and
${\cal X}_{3/2}$
contains the generators corresponding to the parameters $\xi_m$.
Obviously ${\cal X}_{3/2}$ constitutes a solvable algebra of
dimension $r$. Also
${\cal X}_0$ contains a solvable algebra (of dimension
$q\!+\!2$). This follows directly from the observation that the
subalgebra consisting of the generators associated with the
parameters $\xi^\mu$ and $A_{\mu\nu}$ constitute $so(q\!+\!1,1)$,
which, by its Iwasawa decomposition, contains a solvable subalgebra of
dimension $q\!+\!1$
and rank 1 (for $q\geq 0$; for $q=-1$ the algebra is empty, so
that the rank is 0).\footnote{The
action of $SO(q\!+\!1,1)$ on the spinor coordinates follows from
the explicit terms in (\ref{htrans}) proportional to $\xi_\mu$ and
the generators (\ref{cover}) contained in $A_{mn}$ corresponding
to the cover of
$SO(q\!+\!1)$. The additional generators in $A_{mn}$
corresponding to \eqn{mpgc} are compact; they commute with
$SO(q\!+\!1,1)$ and have no bearing on the solvable
subalgebra of ${\cal X}_0$.}
Indeed, the subspace of the special real manifold corresponding
to $h^m=0$ and $h^1$ fixed and non-zero, corresponds precisely to
the coset space $SO(q\!+\!1,1)/SO(q\!+\!1)$.

The complete solvable transitive group of motions thus consists of the
transformations \eqn{htrans} corresponding to the parameters
$\xi _a$, combined with (for $q\geq 0$)
\begin{equation}
A_{\mu \nu }=4\delta _{3[\mu }\xi _{\nu ]}\ ;\qquad A_{mn}= \left( \gamma
_{[3}\gamma _{\mu ]}\right) _{mn}\xi _\mu \ , \end{equation}
where 3 denotes some arbitrary direction in the space of vectors
labeled by indices $\mu $.\vspace{.6cm}

Let us now discuss the implications of our results for the
homogeneous special real spaces with a transitive isometry group
that constitutes an invariance of the polynomial $C(h)$, and thus of
the corresponding $N\!=\!2$ supergravity theory in five
space-time dimensions. These spaces are classified in terms of
the polynomials $C(h)$, as given in (\ref{genC}). The rank of these spaces
is equal to 1 or 2, because the Cartan subalgebra of the solvable algebra
consists of the transformations $\xi _2$, and the Cartan subalgebra of the
solvable algebra corresponding to $SO(q+1,1)/SO(q+1)$. The rank-1 spaces
have $q=-1$ and the corresponding expression for $C(h)$ is
\begin{equation}
L(-1,r):\qquad C(h) = 3\,h^2\,\big( h^1\, h^2
-\big(h^m\big)^2 \big) \ .
\end{equation}
Their solvable algebra is that of
\begin{equation}
L(-1,r):\qquad {SO(r\!+\!1,1)\over SO(r\!+\!1)}\ , \qquad (n=r+2)
\end{equation}
and we therefore identify them with these spaces. They are thus
symmetric and were exhibited in the context of
$d\!=\!5$ supergravity in \cite{GuSiTo2}.
A simple counting argument shows, however, that not all the ${1\over
2}(r+1)(r+2)$ symmetries of this space correspond to
invariances of the cubic polynomial $C(h)$, as there are only $r$
invariances associated with ${\cal X}_{3/2}$ and ${1\over2}r(r-1)
+ 1$ with ${\cal X}_0$ (corresponding to $A_{mn}\sim SO(r)$ and
$\xi_2$, respectively). Indeed, explicit calculations \cite{dWVPVan} show
that the missing $r$ isometries do {\em not} correspond to linear
transformations of the coordinates $h^A$. The case $r=0$ is an
exception in this respect, as all isometries of the real manifold
coincide with the invariances of $C(h)$. The non-linear transformations of
$h$ are not full invariances of the full $d=5$ supergravity action (only
of the scalar part \eqn{sigma}), and the lower dimensional actions do
therefore not exhibit these invariances. This is the reason why
the K\"ahler and quaternionic spaces resulting from the \cmap\
and the \crmap\ applied to $L(-1,r)$ are in general {\em not}
symmetric, with
the exception of the spaces corresponding to $L(-1,0)$ \footnote{In
\cite{GuSiTo2} it was assumed that the space remains
symmetric after reduction; therefore the corresponding K\"ahler
spaces were incorrectly identified with
the minimal couplings of $d=4,\ N=2$ supergravity.}.
Their quaternionic counterparts are missing in the classification
of homogeneous spaces in \cite{Aleks} and the corresponding
K\"ahler spaces are therefore  also missing in \cite{Cecotti}.

The rank-2 spaces with $q=0$ are special, because
$C(h)$ factorizes in certain cases (corresponding to the
symmetric spaces where either $P$ or $\dot P$ vanishes),
\begin{eqnarray}
L(0,P,\dot P):\quad C(h) &=&- 3\Big\{ h^1\,
\big(h^2+h^3\big)\,\big(h^2-h^3\big)\\
&& \qquad +\big(h^2-h^3\big)\,
\big(h^x\big)^2  +\big(h^2+h^3\big)\,
\big(h^{\dot x}\big)^2\Big\} \ ,    \nonumber
\end{eqnarray}
where we have decomposed the indices $m$ into $P$ indices $x$ and
$\dot P$ indices $\dot x$, as explained in the preceding
sections, with $n=3+P+\dot P$. The quaternionic and K\"ahler
spaces corresponding to $L(0,P,\dot P)$ were called $W(P,\dot P)$
and $K(P,\dot P)$ in \cite{Aleks} and \cite{Cecotti},
respectively. We shall denote the real spaces by $Y(P,\dot P)$.

The rank-2 spaces corresponding to $L(q,P)$ with $q>0$ have a
rank-4 quaternionic extension and a rank-3
K\"ahler extension, which were denoted by $V(P,q)$ and $H(P,q)$ in
\cite{Aleks} and \cite{Cecotti}, respectively. We shall denote
the corresponding real spaces by $X(P,q)$.

According to the classification of \cite{Aleks} and
\cite{Cecotti}, for $q=4m\geq 4$ one has precisely one
quaternionic and one K\"ahler space of given (allowed) dimension.
However, the
existence of inequivalent real representations of the Clifford algebra for
$q=4m$ implies the existence of inequivalent real, K\"ahler and
quaternionic spaces corresponding to $L(4m, P,\dot P)$.  We
already encountered an example of the same phenomenon for $q=0$.

As follows from the above arguments quaternionic spaces originating from
special real spaces via special K\"ahler spaces have rank 3 or 4.  But as
mentioned before, these do not constitute all possible homogeneous
quaternionic spaces.  In fact, we know rank-2 symmetric quaternionic
spaces, which originate from special K\"ahler spaces, but not from real
spaces, and rank-1 symmetric quaternionic spaces that also have no
K\"ahler origin.  We summarize these in table~\ref{homnonsp}.
\begin{table}[tb]
\begin{center}
\begin{tabular}{||c|c|c||c|c||}
\hline
real & K\"ahler & quaternionic & $n\!+\!1$& $R$ \\
\hline &&&&\\[-3mm]
&   &   SG     &$0$&0\\[2mm]
&   & $\frac{USp(2n+2,2)}{USp(2n+2)\otimes SU(2)} $
    &$n+1\geq 0$&1 \\[2mm]
&  SG       &$\frac{U(1,2)}{U(1)\otimes U(2)} $
    & 1&1\\[2mm]
&$\frac{U(n,1)}{U(n)\otimes U(1)}$
    &$\frac{U(n+1,2)}{U(n+1)\otimes U(2)} $ & $n+1\geq 2$
    &2    \\[2mm]
SG  & $\frac{SU(1,1)}{U(1)}$
    &$\frac{G_{2(+2)}}{SU(2)\otimes SU(2)}$ &2 &2\\[2mm]
\hline
\end{tabular}
\end{center}
\caption{Normal quaternionic spaces with rank $R\leq 2$ and
quaternionic dimension $n\!+\!1$ and the corresponding special
real and K\"ahler spaces (whenever they exist).}
\label{homnonsp}
\end{table}
\begin{table}[tb]
\begin{center}
\begin{tabular}{||l||c|c|c||c|c||}\hline
$C(h)$&real & K\"ahler & quaternionic & $R$& \\
\hline&&&&&\\[-3mm]
$L(-1,m-1)$&$\frac{SO(m,1)}{SO(m)}$& $\star$ & $\star$ &3&$m\geq 2$\\[2mm]
$L(-1,0)$&$SO(1,1)$&$\left[\frac{SU(1,1)}{U(1)}\right]^2$&$\frac{SO(3,4)}{(
S U ( 2 ) ) ^ 3 } $ &3&\\[2mm]
\hline&&&&&\\[-3mm]
$L(0,P,\dot P)$&$Y(P,\dot P)$&$K(P,\dot P)$&$W(P,\dot P)$&4&$P,\dot P\geq
0$\\[2mm] $L(q,P)$&$X(P,q)$&$H(P,q)$&$V(P,q)$&4&$P,q\geq 1$\\[2mm]
$L(4m,P,\dot P)$& $\star$ & $\star$& $\star$& 4&$m,P,\dot P\geq 1$\\[2mm]
\hline \end{tabular}
\end{center}
\caption{Homogeneous special real spaces with corresponding
K\"ahler and quaternionic spaces. Those that were discussed for the
first time in this paper are indicated by a $\star$. $R$ is the rank of the
quaternionic space. } \label{homsp}
\end{table}
The corresponding K\"ahler and real spaces have real and complex
dimension $n-1$ and $n$, and their rank is equal to $R-2$ and
$R-1$, respectively. Because of the low rank, only a real space
with zero rank can occur (which necessarily has zero dimension).
This corresponds precisely to the pure $N\!=\!2$ supergravity
theory in five space-time dimensions. In the table, this case is
represented by "SG". A similar situation occurs for $R=1$ and
$R=0$, where
the only possibility for a special K\"ahler and quaternionic space
corresponds to pure supergravity in four and three dimensions,
respectively. Hence none of the spaces discussed in the table are
related to the spaces classified in this paper. Observe that all
spaces in table~\ref{homnonsp} are symmetric. Together with the
homogeneous spaces resulting from the analysis of this paper,
which are summarized in table~\ref{homsp},
they constitute all the homogeneous quaternionic and special
K\"ahler spaces that are known.
A proof that this list contains all the symmetric special K\"ahler spaces is
given in \cite{CremVP}. The symmetric rank-4 quaternionic
spaces and their related special real and K\"ahler spaces correspond
to $L(0,P),\ L(1,1),\ L(2,1),\ L(4,1)$, and $L(8,1)$.

These tables show a remarkable pattern. We have the pure $N=2$ supergravity
theory in 3 dimensions ("the empty quaternionic space") and the minimal
couplings~: the quaternionic projective spaces. Then the remaining rank-1
quaternionic symmetric space is the one originating from pure $d=4$
supergravity ("the empty special K\"ahler space"). The minimal couplings of
vector multiplets in $d=4,\ N=2$ supergravity, the complex projective
spaces, are the origin of an infinite series of rank-2 quaternionic
spaces. The remaining rank-2 quaternionic space originates from pure $d=5$
supergravity ("empty real space"), while the real projective
spaces are the origin of an infinite series of rank-3 homogeneous
quaternionic spaces (as discussed before, the reduction does not
preserve the property that the space be symmetric). Seeing the
ensuing pattern, it looks as if the remaining rank-3 quaternionic
space should arise from the reduction of pure
$d=6$ supergravity. The rank-4 quaternionic spaces would then find their
origin in matter coupled $d=6$ supergravity. This is then also the last
step, because $d=6$ is the largest space-time dimension in which a
supergravity theory can exist with 8 independent supersymmetries
(corresponding to a $d=6$ spinor).  These $d=6$ couplings would then be
characterized by
the possible real realizations of positive-definite Clifford
algebras (while $L(-1,0)$ corresponds to the "empty Clifford algebra").
This is in accord with a conjecture in \cite{Romans}
(cf.  eq.(5.6)) where
$d=6,\ N=2$ tensor and vector multiplets are
incorporated in the field strength
\begin{equation}
F_{abc }^\mu =3\partial _{[a}A_{bc]}^\mu +\left( A^m\right) _{[a}
\left(\gamma^\mu \right) _{mn}\left( F^n\right) _{bc]},
\end{equation}
which leads to a coupling of $q+1$ tensor multiplets (with tensor
field $A^\mu _{ab}$ and field strength $F^\mu _{abc}$) to $r$
vector multiplets (with vectors $A^m_a$ and field strengths
$A^m_{ab}$).

In this paper we
presented a complete classification of the special real homogeneous
spaces with a transitive group of motions that leaves the polynomial
$C(h)$, and thus the corresponding $d=5$ supergravity theory, invariant.
Therefore we also obtained the corresponding classification for
the homogeneous special K\"ahler and quaternionic spaces
that are in the image of the \cmap\ and
the \crmap.  However, we expect that the tables~\ref{homnonsp} and
\ref{homsp} in fact comprise
all possible homogeneous quaternionic spaces.
This result should still follow from the analysis of
\cite{Aleks}, and we believe that the absence in \cite{Aleks} of the spaces
indicated by a $\star$ in table~\ref{homsp} is merely due to a
calculational error.  The nice pattern described above lends support to our
conjecture that the classification of homogeneous quaternionic spaces is
now complete, as the new spaces exhibited above are precisely needed for
completing the overall picture.
\vspace{0.6cm}

\noindent{\large \bf Acknowledgments}\vspace{0.3cm}

We thank S.~Cecotti, R.~Coquereaux, A.S.~Galperin, V.I.~Ogievetsky,
W.~Troost and F.~Vanderseypen for valuable discussions.

\end{document}